\documentclass[10pt,a4paper]{article}
\usepackage[utf8]{inputenc}
\usepackage[english]{babel}

\usepackage{amsmath}
\usepackage{amsfonts}
\usepackage{amssymb}

\usepackage[colorlinks,citecolor=blue,urlcolor=blue,linkcolor=blue]{hyperref}

\usepackage[left=2cm,right=2cm,top=2cm,bottom=2cm]{geometry}

\usepackage{authblk}

\usepackage{graphicx}
\usepackage{subcaption}

\usepackage{feynmp}
\newcommand{\pd}{\partial}
\newcommand{\abs}[1]{\left\lvert #1 \right\rvert }
\newcommand{\okappa}[1]{\mathcal{O}\left( \kappa^#1 \right)}

\newcommand{\rh}{r_\text{h}}

\newcommand{\rmd}{{\rm d}}

\title{Black Holes in Einstein-scalar-Gauss-Bonnet model probed with scattering amplitudes}
\author[1]{\href{https://orcid.org/0000-0001-7099-0861}{Boris Latosh} \thanks{ \href{mailto:latosh.boris@ibs.re.kr}{latosh.boris@ibs.re.kr} }}
\author[1]{\href{https://orcid.org/0000-0002-9950-0069}{Miok Park} \thanks{ \href{mailto:miokpark76@ibs.re.kr}{miokpark76@ibs.re.kr} }}
\affil[1]{Particle Theory  and Cosmology Group, Center for Theoretical Physics of the Universe, Institute for Basic Science (IBS), Daejeon, 34126, South Korea}
\date{CTPU-PTC-23-46}

\usepackage{feynmp}
\begin{document}

\maketitle

\begin{abstract}
We examined the quantum properties of scalar-tensor gravity with a coupling to the Gauss-Bonnet term in the low energy limit, exploring both linear and quadratic couplings. We calculated the leading-order corrections to the non-relativistic one-body gravitational potential and the metric by studying the gravitational field of a point-like scalar particle. We studied light-like scattering and compared it with the classical theory. We found that the non-minimal coupling does not contribute to the small-angle scattering for the quadratic coupling but does in the case of linear coupling. The results provide an opportunity to constrain the linear non-minimal coupling to the Gauss-Bonnet term with forthcoming observational data.
\end{abstract}

\section{Introduction}

Creating an ultraviolet complete quantum theory of gravity is one of the most challenging problems of contemporary physics. One of the major obstacles is the non-renormalizability of general relativity \cite{tHooft:1974toh,Goroff:1985th,Stelle:1976gc}. In contrast with renormalisable theories, general relativity generates new operators at each level of perturbation theory. One can subtract ultraviolet divergences in any amplitude at any order of the perturbation theory. However, the subtraction requires initial data to normalise the finite part of the expression. Since the theory generates new operators at every order of perturbation theory, it also needs a new set of initial data at every new order, and the theory loses predictability.

Perturbative quantum gravity aligns with the effective field theory paradigm, which does not aim to provide a complete ultraviolet theory but instead focuses on applicability in low-energy regimes \cite{Donoghue:1994dn,Burgess:2003jk,Vanhove:2021zel}. Perturbative quantum gravity is a quantum theory describing small metric perturbations propagating around flat spacetime. The theory remains applicable if metric perturbations are small, which constrain the theory to the sub-Planck region. The theory remains non-renormalisable and requires new initial data at each new perturbation theory level. The lack of renormalizability within the effective field theory is due to the need for more information on its ultraviolet extension since each new perturbative correction extends the theory in the ultraviolet region and requires additional data.

The perturbative approach to quantum gravity offers a powerful tool to calculate scattering amplitudes involving gravitons using the standard tools of quantum field theory. Recent advancements in scattering amplitude calculations have further validated the applicability of perturbative quantum gravity \cite{Elvang:2013cua,Elvang:2015rqa,Arkani-Hamed:2017jhn,Travaglini:2022uwo,Bjerrum-Bohr:2022blt,Badger:2023eqz}. These methods use causality and unitarity relations to recover amplitudes without direct reliance on Feynman rules. The corresponding results exist within the effective theory paradigm, confirming perturbative quantum gravity calculations. Despite the active development of methods for scattering amplitude calculations, this paper uses the widely adopted standard methods based on Feynman rules.

The low-energy limit is a valuable tool for studying perturbative quantum gravity. Its key feature is the decoupling of scales, which allows one to study the low-energy behaviour independently from possible ultraviolet extensions. The low-energy limit takes place when the spatial momenta of all particles approach zero. Consequently, the particle interaction energy also approaches zero. The BPHZ theorem \cite{Bogoliubov:1957gp,Hepp:1966eg,Zimmermann:1969jj,Lam:1973qa} governs the theory behaviour in this limit. According to the theorem, ultraviolet divergences are always multiplied by operators, which are analytic functions of external momenta. Consequently, all terms with ultraviolet divergences remain bounded in the low energy limit and do not grow. The ultraviolet finite terms, in contrast, are not constrained by the theorem and can contain non-local operators, which are non-analytic functions of external momenta. Such non-analytic operators are not bounded and grow uncontrollably in the low energy limit. Consequently, in the low energy limit, the leading order contribution is entirely free from ultraviolet divergences and is not influenced by renormalisation. Terms that do contain ultraviolet divergences and are influenced by a renormalisation procedure are suppressed by kinematics and can be neglected.

The low energy limit in perturbative quantum gravity is widely used. In the following sections, we discuss its implementation in great detail. Previous research provide extensively review of this topic \cite{Donoghue:2001qc, Burgess:2003jk, Elvang:2013cua, Elvang:2015rqa, Bjerrum-Bohr:2021vuf, Vanhove:2021zel, Travaglini:2022uwo, Bjerrum-Bohr:2022blt, Badger:2023eqz}.

This paper examines the low-energy properties of specific scalar-tensor gravity models with coupling to the Gauss-Bonnet term within perturbative quantum gravity \cite{Sannan:1986tz,Fujii:1996td,Shojai:1998jhg,Shojai:2000us,Latosh:2020jyq,Arbuzov:2020pgp,Latosh:2021usy}. We study the influence of the non-minimal coupling on the external gravitational field of massive bodies and investigate the opportunity to verify this influence by small-angle light scattering. Studying such models is important since the existing apparatus allows us to verify models of gravity directly and test the limits of general relativity's applicability.

Another particular feature of such models is the evasion of the no-hair theorem. The no-hair theorem ensures that a black hole has only three macroscopic parameters: mass, angular momentum, and electric charge \cite{Nojiri:2010wj,Berti:2015itd}. The theorem holds in general relativity and some modified gravity models. Scalar-tensor models with a coupling to the Gauss-Bonnet term present an exceptional case when the theorem fails \cite{Antoniou:2017acq,Lee:2018zym,Papageorgiou:2022umj,Nojiri:2023jtf}. The theory allows black holes to form a non-trivial external scalar field, i.e. the scalar hair. No analytic solution for such black holes was found, but a few numerical solutions have been constructed \cite{Papageorgiou:2022umj}.

The existence of black holes with scalar hair holds immense significance as they exhibit a completely different phenomenology compared to black holes within general relativity. The external gravitational field of a black hole with scalar hair deviates from the general relativity case, providing a unique opportunity to test the model. Notably, the scalar hair can influence the motion of light around such a black hole, forming a shadow distinct from those described by general relativity. This intriguing possibility can be tested with contemporary apparatus, such as the Event Horizon Telescope collaboration \cite{EventHorizonTelescope:2019dse}, sparking anticipation for potential groundbreaking discoveries.

The Gauss-Bonnet term itself occupies a special place within modified gravity as well. The term involves higher curvature terms, but it is a complete derivative in $d = 4$ \cite{Lovelock:1971yv,Lovelock:1972vz,Horndeski:1974wa,Horndeski:1976gi,Kobayashi:2011nu}:
\begin{align}
  \mathcal{G} = R^2 - 4R_{\mu\nu}R^{\mu\nu} + R_{\mu\nu\alpha\beta}R^{\mu\nu\alpha\beta} .
\end{align}
When the term couples to a scalar field, it is no longer a complete derivative and contributes to the classical field equations. Despite the term involving higher derivatives, it does not introduce higher derivatives to the field equations \cite{Horndeski:1974wa,Horndeski:1976gi,Kobayashi:2011nu}. Lastly, the presence of the Gauss-Bonnet term is typical for string models, so it appears essential for a consistent ultraviolet gravity extension.

This paper examines two cases of non-minimal coupling between a scalar field and the Gauss-Bonnet term. Namely, we investigate couplings that are linear and quadratic in the scalar field. We study the gravitational field of a scalar particle with this non-minimal coupling and calculate the leading-order corrections to the two-body gravitational potential and metric. We use these results to determine whether small-angle light scattering can constrain such corrections.

The gravitational field of a point-like particle well approximates the leading contribution to the gravitational field of a massive body, as a distant observer can neglect the influence of the body's internal structure on its gravitational field. In the classical case, the post-Newtonian expansion of the gravitational field of a point particle matches the expansion of the Schwarzschild metric \cite{will2018theory,Will:2014kxa,Straumann:2013spu}. Perturbative quantum gravity shows the same behaviour. Expansions of gravitational field for massive scalar, massive fermion, and massive charged fermion match expansions of the Schwarzschild, Kerr, and Kerr-Newman metrics \cite{Donoghue:1994dn,Donoghue:2001qc,Bjerrum-Bohr:2002fji}.

We employ the same approach for the discussed models. We calculate the leading corrections to the gravitation field of a particle generated by the non-minimal coupling. We recover the two-body gravitational potential and the metric describing the gravitation field of a single particle. After this, we associate this metric with the leading contribution to the metric of a massive body.

The small-angle light scattering provides an opportunity to verify the theory. The influence of the non-minimal coupling is expected to be suppressed by the new mass scale and fall rapidly with distance. Nonetheless, we show that it contributes to the small angle scattering in the linear coupling case. In turn, such corrections can be constrained empirically with black hole data. Since black holes are massive compact objects, the non-minimal coupling will also influence their gravitational field. In turn, one can use the weak lensing data or the data on black hole shadows to constrain the magnitude of such corrections.

The paper is organised as follows. Section \ref{Section_black_hole_metric} briefly discusses the Einstein-scalar-Gauss-Bonnet theory and its numerical black hole solutions. Section \ref{Section_perturbative_quantum_gravity} discusses the methodology of perturbative quantum gravity and the low energy limit. Section \ref{Section_metric_reconstruction} applies the perturbative quantum gravity to the discussed models. We calculate the leading corrections to the two-body potential and the metric describing the gravitation field of a particle with such non-minimal coupling. Section \ref{Section_classical_scattering} discusses the classical light scattering on a black hole with a focus on small-angle scattering. In Section \ref{Section_quantum_scattering}, we calculate the leading quantum corrections to that scattering generated by the Gauss-Bonnet term. We show that some models can meaningfully contribute to the small angel scattering. We present our conclusions in Section \ref{Section_conclusion}.

\section{Numerical Black Hole Solutions}\label{Section_black_hole_metric}

We begin with a scalar-tensor model featuring non-minimal coupling to the Gauss-Bonnet term, defined by the following action:
\begin{align}\label{the_action}
  S_{\textrm{grav.}} &= \int \rmd^4 x \sqrt{-g} \left[- \cfrac{1}{16\,\pi\,G_\text{N}}\,R +\cfrac{1}{2}\, g^{\mu\nu}\, \nabla_\mu \varphi \, \nabla_\nu \varphi - f(\varphi)\, \mathcal{G} \right] .
\end{align}
$G_\text{N}$ denotes the Newton constant, $f(\varphi)$ represents the function describing the non-minimal coupling, and $\mathcal{G}$ is the Gauss-Bonnet term:
\begin{align}
  \mathcal{G} = R^2 - 4 \, R_{\mu\nu}^2 + R_{\mu\nu\alpha\beta}^2 \,.
\end{align}
This action produces the following field equations:
\begin{align}\label{the_field_equations_covariant_form}
  \begin{split}
    R_{\mu\nu} - \cfrac12\, R\, g_{\mu\nu} & = 8\,\pi\,G_\text{N} \left[ \nabla_\mu \varphi \, \nabla_\nu \varphi - \cfrac12\, g_{\mu\nu} \, \left(\nabla\varphi\right)^2 - \left(g_{\mu\alpha}g_{\nu\beta} + g_{\mu\beta} g_{\nu\alpha}\right)\, P^{\alpha\rho\beta\sigma} \, \nabla_\rho\nabla_\sigma f(\varphi)\right], \\
    \square\, \varphi + \dot{f}(\varphi) \, \mathcal{G} & = 0.
  \end{split}
\end{align}
Throughout this paper, $\dot{f} = \pd f / \pd \varphi$ denotes the derivative with respect to $\varphi$, and $P^{\mu\nu\alpha\beta}$ is the Hodge dual of the Riemann tensor\footnote{The numerical factor used for this definition varies in different sources.}:
\begin{align}
  \begin{split}
    P^{\mu\nu\alpha\beta} &= \eta^{\mu\nu\mu'\nu'} \, \eta^{\alpha\beta\alpha'\beta'} \,R_{\mu'\nu'\alpha'\beta'}\,, \\
    \eta^{\mu\nu\alpha\beta} &= \cfrac{ \varepsilon^{\mu\nu\alpha\beta}}{\sqrt{-g}} \,.
  \end{split}
\end{align}
Although it includes higher-derivative terms, this model belongs to the Horndeski class, yielding second-order field equations \cite{Horndeski:1974wa,Horndeski:1976gi,Kobayashi:2011nu}.

Following \cite{Papageorgiou:2022umj}, we describe the construction of numerical black hole solutions in the remainder of this section. For consistency, this section will use the metric signature $(-+++)$ and units where $16\,\pi\, G_\text{N}=1$. We use the standard ansatz for a spherically symmetric metric:
\begin{align}
 \rmd s^2 = - A(r) \rmd t^2 + \frac{1}{B(r)}\,\rmd r^2 + r^2 \, (\rmd \theta^2 + \sin^2 \theta \rmd \phi^2) \,. \label{eq:ansatz}
\end{align}
It brings the field equations to the following form:
\begin{align}
  & \cfrac{1}{r^2} \left[ 1 - \frac{1}{B} + r\, \frac{B'}{B} \right] + \frac{1}{4}\,\varphi'^2 - \frac{2}{r^2}\left[ \frac{B'}{B} \,(3\,B-1) f' + 2\,(B-1)\,f'' \right] = 0 ,\label{eq:tt} \\
  & \cfrac{1}{r^2} \left[ 1 - \frac{1}{B} + r\, \frac{A'}{A} \right] - \frac{1}{4}\,\varphi'^2 - \frac{2}{r^2} \, \frac{A'}{A}\,(3\,B-1) f' = 0, \hspace{150pt} \label{eq:rr} \\
  & \cfrac{1}{2}\left[ \frac{A''}{A} - \frac12\,\left(\frac{A'}{A}\right)^2 + \frac{1}{r}\,\ln'[A\,B] + \frac12\,\frac{A'}{A}\,\frac{B'}{B} \right] +\frac{1}{4}\, \varphi'^2 -\cfrac{1}{r}\left[ B\,\frac{A'}{A}\, f'\,\ln'[A\,B] + 2 \left(B\, \frac{A'}{A} f'\right)' \right] = 0, \label{eq:phiphi} \\
  & \varphi'' + \frac{1}{2}\, \varphi ' \ln' \left[ r^4 \, A\,B\right] + \frac{2}{r^2}\, \dot{f}\,\left[(3\,B-1) \,\frac{A'}{A} \, \frac{B'}{B} - (B-1) \left(\,\left(\frac{A'}{A}\right)^2 - 2 \,\frac{A''}{A}\right)\right] = 0 .\label{eq:scalar}
\end{align}
Here ``\,$'$\,''$=\pd/\pd r$ denotes the derivative by the radial coordinate $r$ and so $f' = \dot{f} \varphi'$. 

\begin{figure}[htbp]
  \begin{minipage}[b]{\linewidth}
    \centering
    \begin{subfigure}[b]{0.49\linewidth}
      \includegraphics[width=\linewidth]{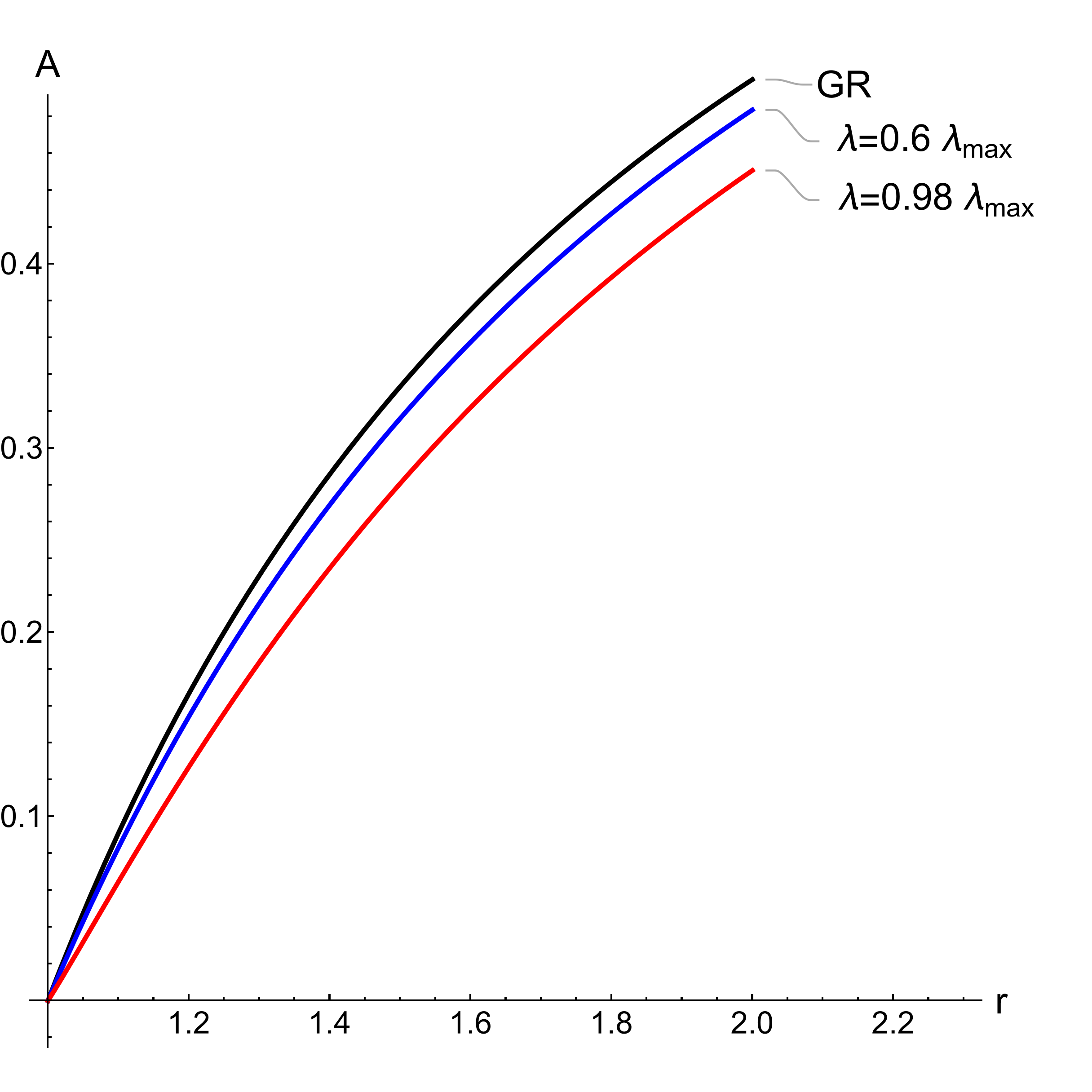}
      \caption{Metric function $A$}
    \end{subfigure}
    \begin{subfigure}[b]{0.49\linewidth}
      \includegraphics[width=\linewidth]{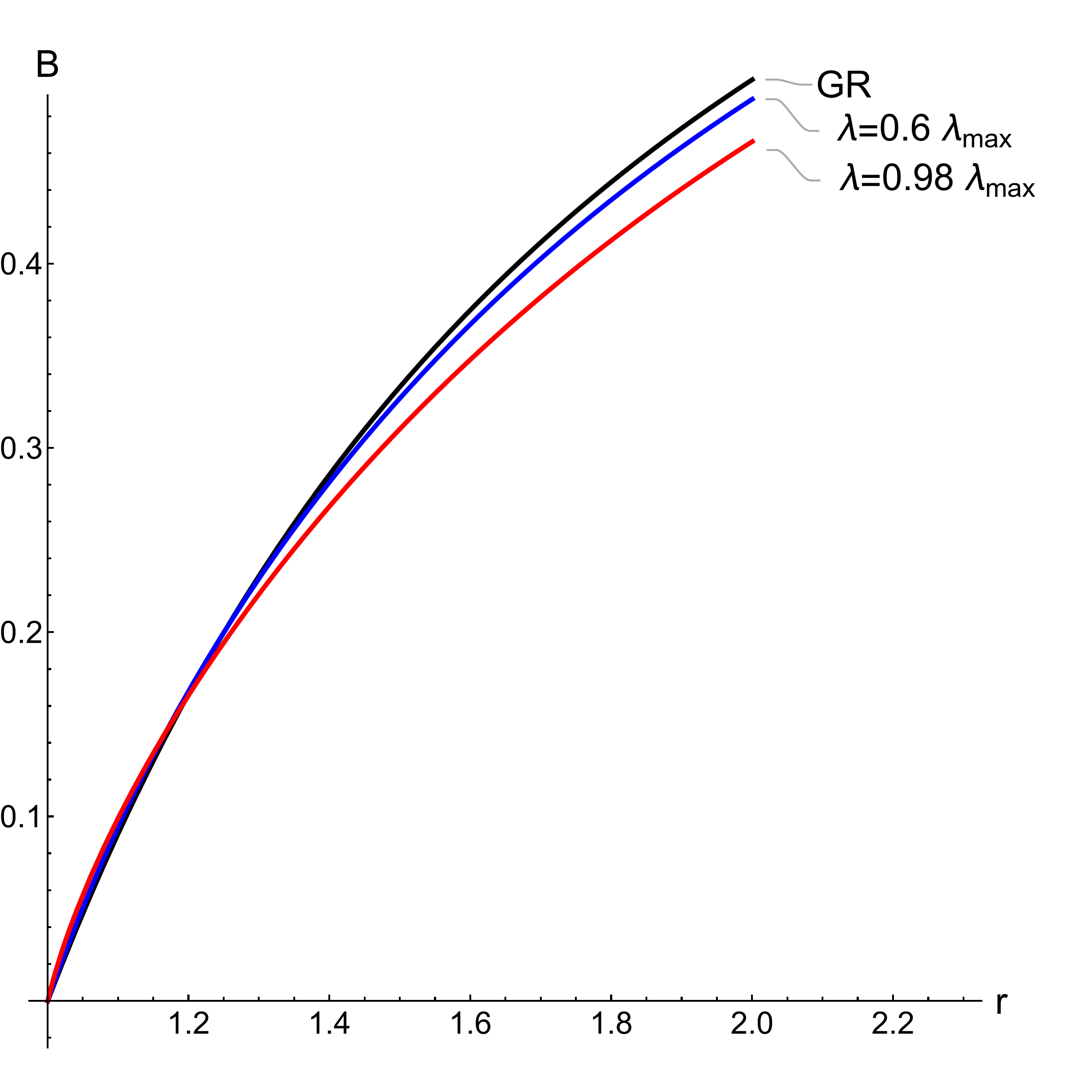}
      \caption{Metric function $B$}
    \end{subfigure}
  \end{minipage}
  \begin{minipage}[b]{\linewidth}
    \centering
    \begin{subfigure}[b]{0.49\linewidth}
      \includegraphics[width=\linewidth]{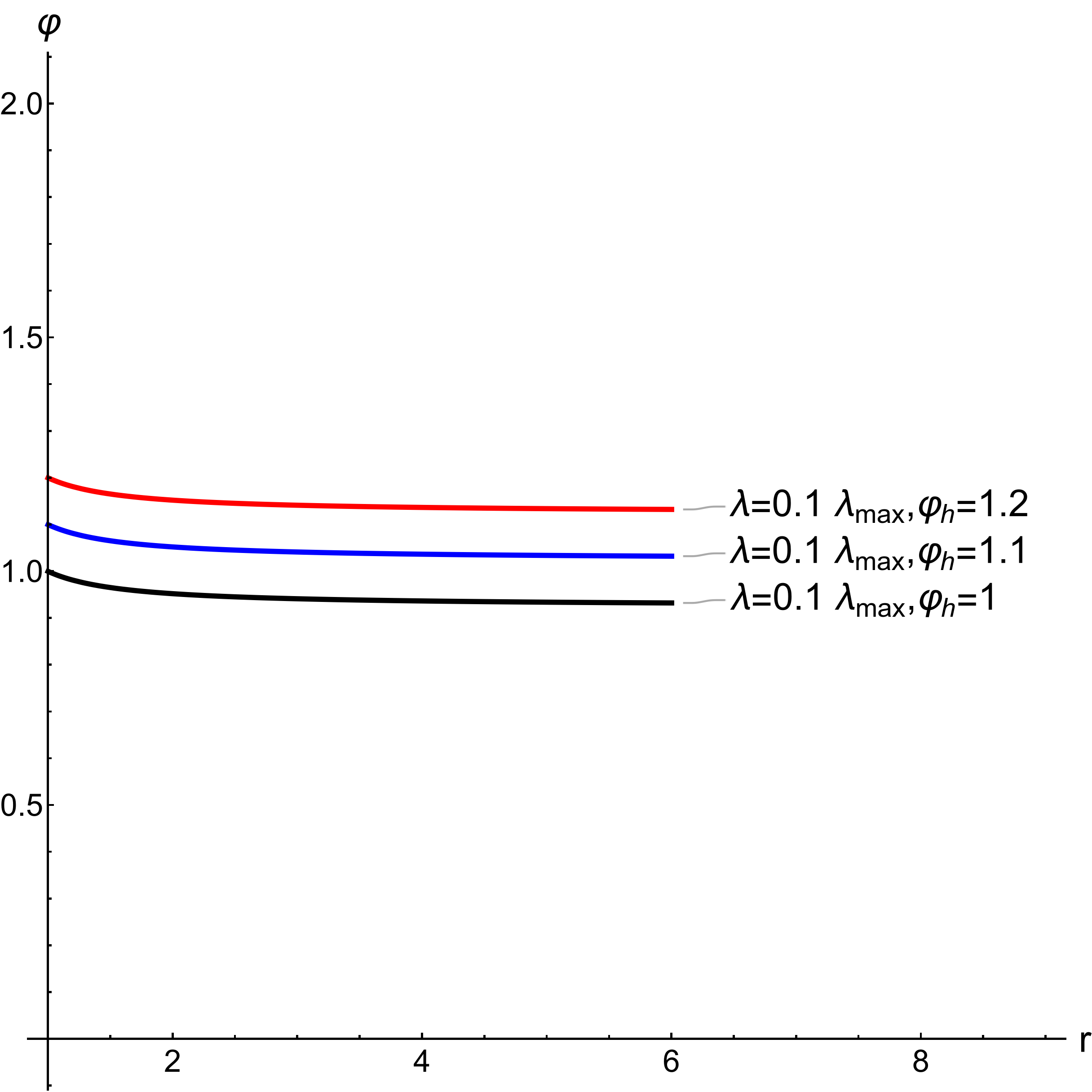}
      \caption{Scalar field $\varphi$}
    \end{subfigure}
    \begin{subfigure}[b]{0.49\linewidth}
      \includegraphics[width=\linewidth]{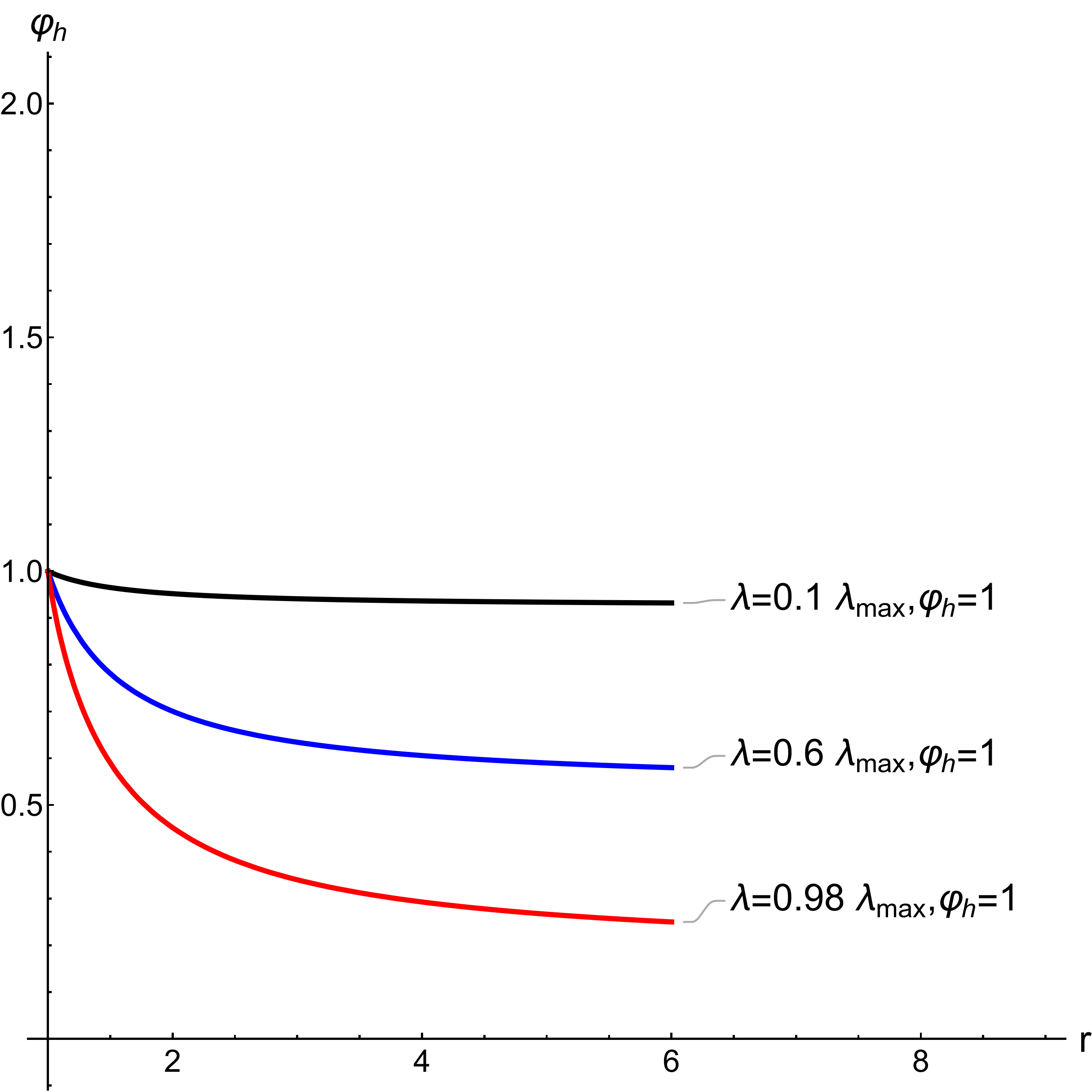}
      \caption{Scalar field $\varphi$}
    \end{subfigure}
  \end{minipage}
  \caption{Functions $A$, $B$, $\varphi$ for the linear coupling. The maximum value $\lambda_\text{max}$ is defined by \eqref{coupling-scalar_constraints} and independent of $\varphi_\text{h}$. In the $\lambda \to 0 $ limit, the metric recovers the Schwarzschild black hole.}
  \label{fig:metric_linear_coupling}
\end{figure}
\begin{figure}[htbp]
  \begin{minipage}[b]{\linewidth}
    \centering
    \begin{subfigure}[b]{0.49\linewidth}
      \includegraphics[width=\linewidth]{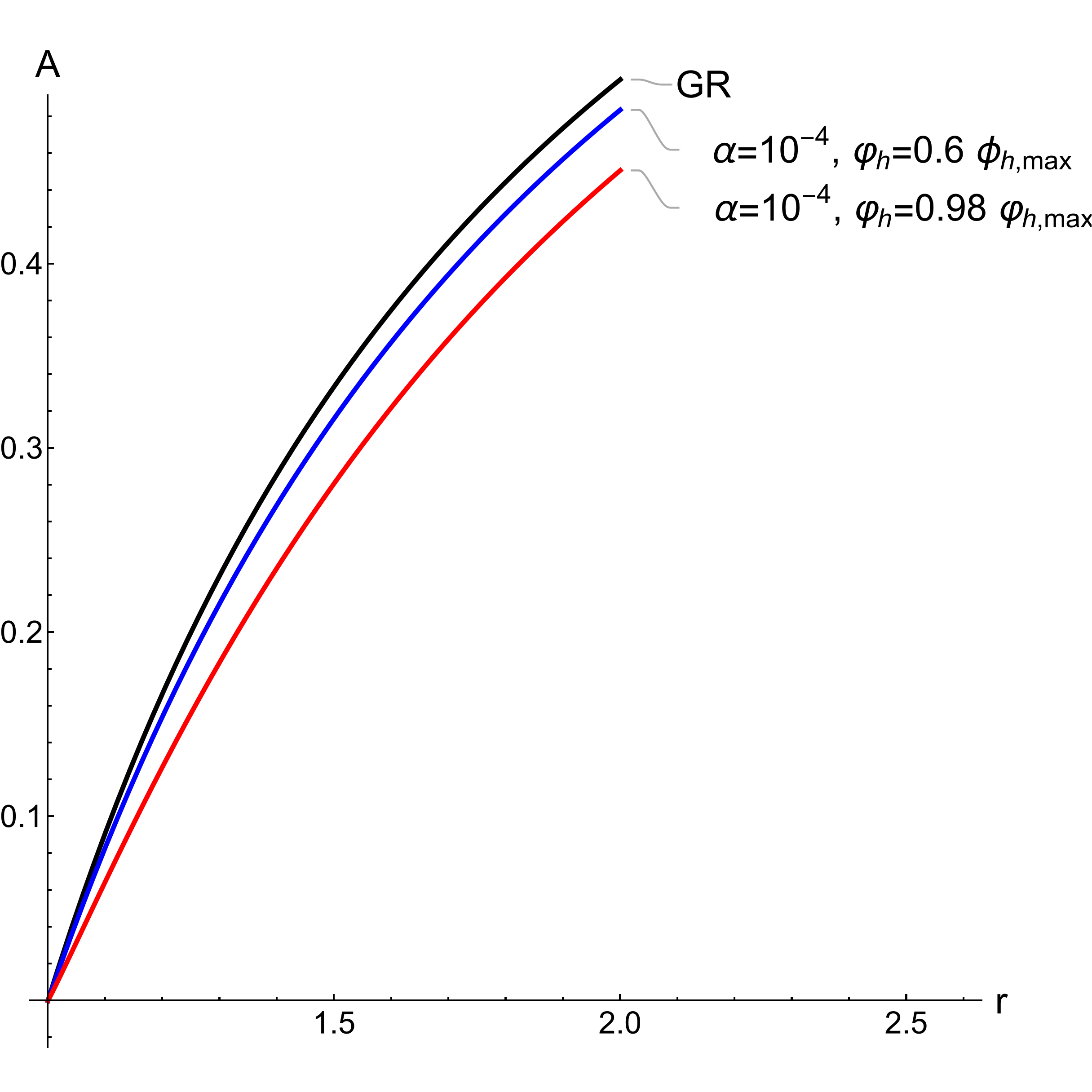}
      \caption{Metric function $A$}
    \end{subfigure}
    \begin{subfigure}[b]{0.49\linewidth}
      \includegraphics[width=\linewidth]{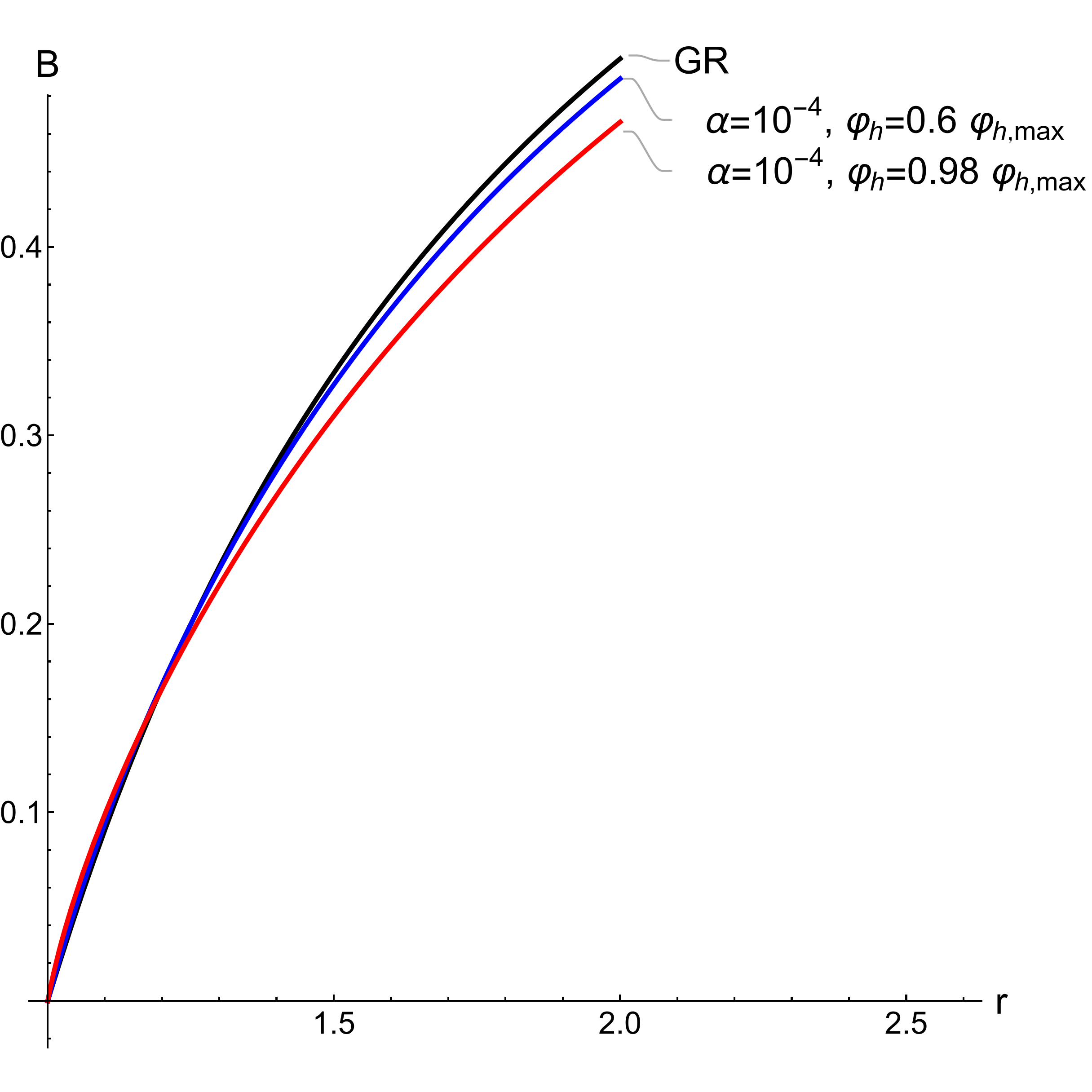}
      \caption{Metric function $B$}
    \end{subfigure}
  \end{minipage}
  \begin{minipage}[b]{\linewidth}
    \centering
    \begin{subfigure}[b]{0.49\linewidth}
      \includegraphics[width=\linewidth]{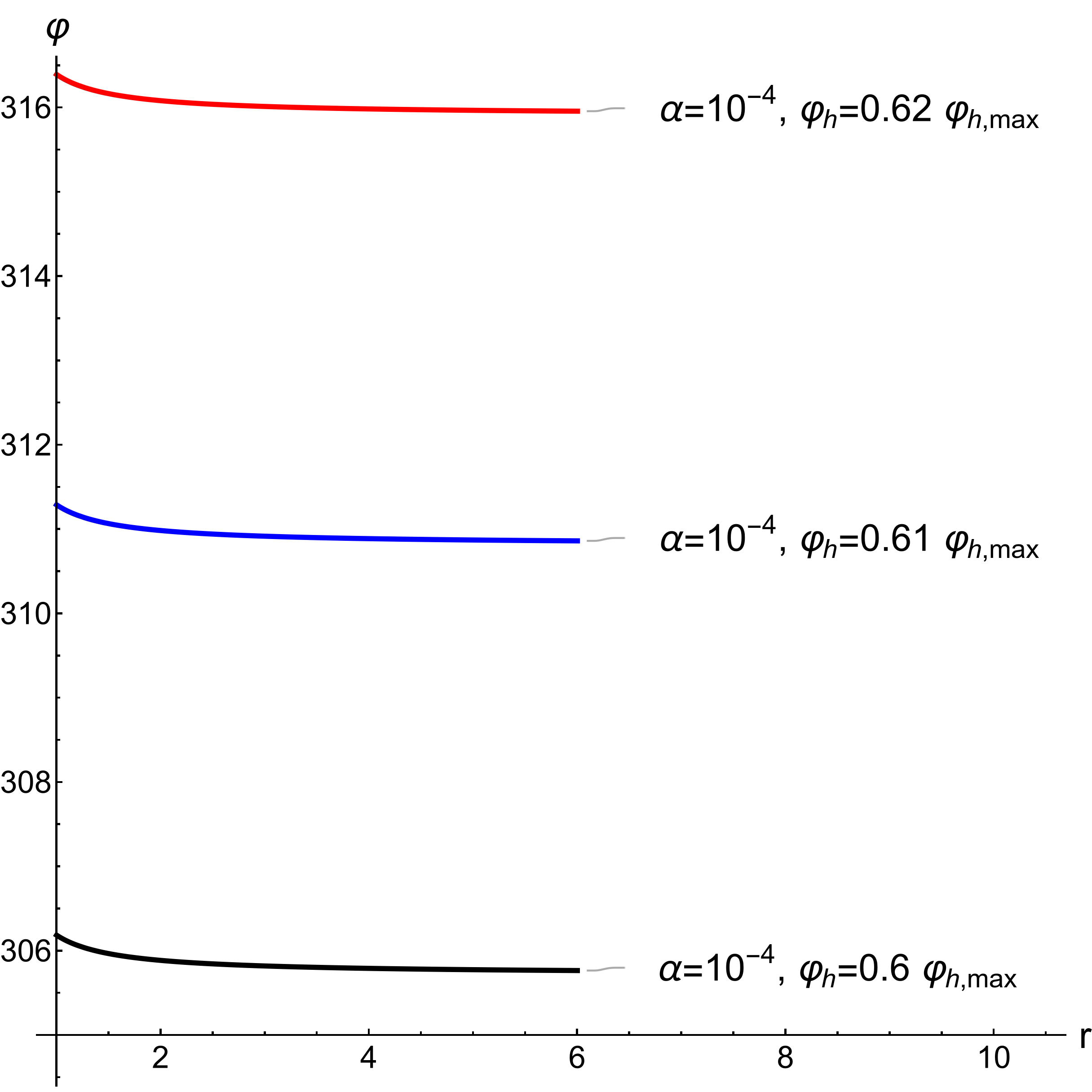}
      \caption{Scalar field $\varphi$}
    \end{subfigure}
    \begin{subfigure}[b]{0.49\linewidth}
      \includegraphics[width=\linewidth]{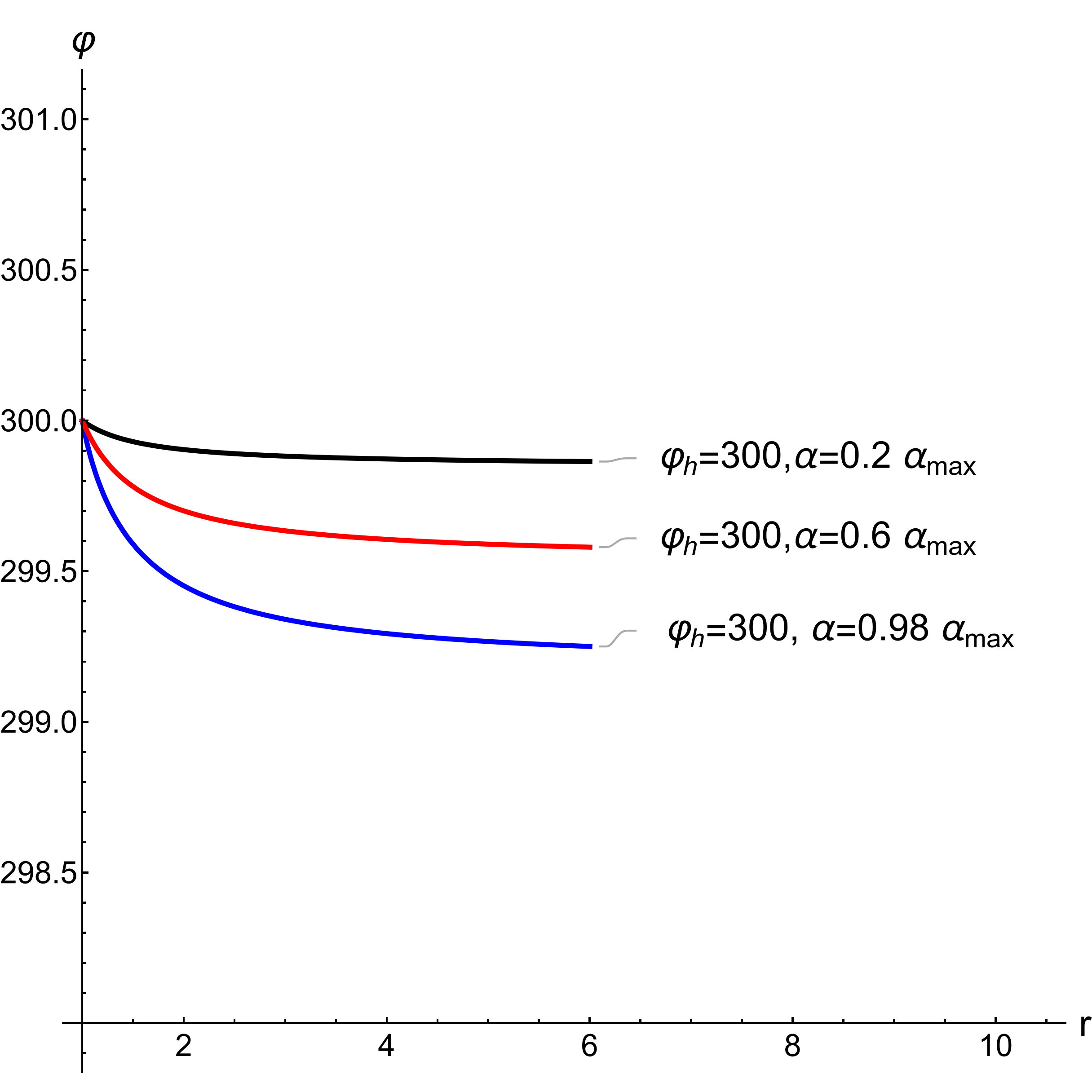}
      \caption{Scalar field $\varphi$}
    \end{subfigure}
  \end{minipage}
  \caption{Functions $A$, $B$, $\varphi$ for the quadratic coupling. The maximum value $\alpha_\text{max}$ is defined by \eqref{coupling-scalar_constraints} and depends on $\varphi_\text{h}$. In $\alpha \to 0 $, the metric recovers the Schwarzschild black hole. When $\alpha \to 0 $, the range of allowed initial values of $\varphi_\text{h}$ expands, and the field profile approaches a constant value defined only by $\varphi_\text{h}$. Similar to the previous case, the metric functions approach those of general relativity.}
  \label{fig:metric_quadratic_coupling}
\end{figure}

The following constraint must be satisfied at the horizon to keep the scalar field and its derivatives finite \cite{Papageorgiou:2022umj,Kanti:1995vq,Antoniou:2017acq,Lee:2018zym}:
\begin{align}
  {\varphi_{\text{h},1}} &= - \cfrac{\rh}{4\,\dot{f}_\text{h}} \left(1 - \sqrt{1 - \frac{96}{\rh^4} \,\dot{f}_\text{h}^2 } \,\right) , \label{eq:dvarphi}\\
  B_\text{h} &= \cfrac{\rh^3}{48 \, \dot{f}_\text{h}^2} \left(1 - \sqrt{1 - \frac{96}{\rh^4} \,\dot{f}_\text{h}^2 } \,\right) ,\label{eq:Bh}
\end{align}
with $\dot{f}_\text{h} = \dot{f}(\varphi_\text{h})$. The boundary conditions near the horizon are determined by two independent parameters $A_\text{h}$ and $\varphi_\text{h}$ with a fixed value of $\rh$. Here $A_{\text{h}}$ is fixed by the requirement of asymptotic flatness at infinity; which requires
\begin{align}\label{eq:INFexp}
  \begin{split}
    A(r) &\overset{r\to\infty}{\rightarrow} 1 ,\\
    B(r) &\overset{r\to\infty}{\rightarrow} 1 ,\\
    \varphi(r) & \overset{r\to\infty}{\rightarrow}\varphi_{\infty}. 
 \end{split}
\end{align}
 Consequently, $\varphi_\text{h}$ becomes the sole free parameter. This allows us to construct a one-parameter family of black holes on $\varphi_\text{h}$ for a given set of values for $\alpha$ and $\lambda$. The value of the scalar field at infinity ($\varphi_\infty$) is determined by the boundary conditions at the horizon. To prevent the divergence of $\varphi''(\rho)$, the value under the square root must not be zero:
\begin{align}
  \left(\dot{f}_\text{h}\right)^{2} < \frac{\rh^4}{96} \,. \label{eq:constraint2}
\end{align}
This condition restricts the values between the coupling constants in $f_\text{h}$ and $\varphi_{\text{h}}$.

This study focuses exclusively on linear and quadratic non-minimal Gauss-Bonnet couplings:
\begin{align}
  f(\varphi) &= \lambda\, \varphi , & f(\varphi) &= \alpha \,\varphi^2 .
\end{align}
For these cases the constraint \eqref{eq:constraint2} takes the following forms:
\begin{align}\label{coupling-scalar_constraints}
  \lvert \lambda \rvert & < \cfrac{\rh^2 }{4 \sqrt{6}} \, , &   \lvert \alpha \varphi_{\text{h}} \rvert < \frac{ \rh^2 }{8 \sqrt{6}}.
\end{align}
These indicate that the horizon value of the scalar field $\varphi_{\text{h}}$ is unconstrained, but the values of $\lambda$ are constrained for the case of linear coupling. For the case of quadratic coupling, the range of $\varphi_{\text{h}}$ is determined when a value of $\alpha$ is given, and vice versa.

To obtain the numerical solution for the hairy black hole, we set the horizon radius to $r_\text{h}=1$ and fix the scalar field value $\varphi_\text{h}$ near the horizon. Numerical solutions for $A$, $B$, and $\varphi$ are presented in Figures \ref{fig:metric_linear_coupling} and \ref{fig:metric_quadratic_coupling}. These figures display solutions with various couplings and boundary conditions.

\section{Perturbative quantum gravity}\label{Section_perturbative_quantum_gravity}

This section briefly reviews perturbative quantum gravity and the usage of the low-energy limit. Existing literature \cite{Burgess:2003jk,Vanhove:2021zel,Latosh:2023zsi} provides a more comprehensive review.

Perturbative quantum gravity is a quantum theory of small metric perturbations $h_{\mu\nu}$ propagating on a flat background. The complete spacetime metric combines the flat background $\eta_{\mu\nu}$ and the perturbations:
\begin{align}\label{perturbative_metric}
  g_{\mu\nu} = \eta_{\mu\nu} + \kappa \, h_{\mu\nu} .
\end{align}
In this expression, $\kappa^2 = 32\pi\, G_\text{N}$ is the gravitational coupling and  $G_\text{N}$ is the Newton constant. 

The metric \eqref{perturbative_metric} contains a finite number of terms in $\kappa$. However, it introduces infinite series in $\kappa$ to the theory. The inverse metric $g^{\mu\nu}$, that enters many expressions, is an infinite series:
\begin{align}
  g^{\mu\nu} = \eta^{\mu\nu} - \kappa\, h^{\mu\nu} + \kappa^2 \, h^\mu{}_\sigma h^{\sigma\nu} + \okappa{3} .
\end{align}
Consequently, quantities like the Christoffel symbols $\Gamma^\alpha_{\mu\nu}$ and the Riemann tensor $R_{\mu\nu}{}^\alpha{}_\beta$, as well as the volume factor $\sqrt{-g}$, all produce infinite series in $\kappa$. Since $\sqrt{-g}$ enters the action for any gravity model, it introduces series in $\kappa$ to all gravity models.

Let us examine the structure of the microscopic action $\mathcal{A}$ of a generic gravity model. The action expands in an infinite series, and each term plays a different role:
\begin{align}\label{Microscopic_Action_Perturbative_Expansion}
  \mathcal{A}[\eta_{\mu\nu} + \kappa h_{\mu\nu}]=\mathcal{A}[\eta_{\mu\nu}] + \left. \frac{\delta\mathcal{A}}{\delta g_{\mu\nu}}\right|_{g=\eta} \, \kappa\, h_{\mu\nu} + \left. \cfrac{\delta^2 \mathcal{A}}{ \delta g_{\mu\nu}\, \delta g_{\alpha\beta} } \right|_{g = \eta} \kappa^2\,h_{\mu\nu} h_{\alpha\beta} + \okappa{3}.
\end{align}
The first term is constant and does not contain metric perturbations. The second term is a linear coupling between perturbations and the first variation of the action. This term vanishes if $\eta_{\mu\nu}$ is a solution of the model classical field equations. Although this condition is not trivial for all gravity models, models considered in this paper fulfil it, so we do not discuss it further. The third term is the kinetic term describing the propagation of small metric perturbation about the flat background. The other terms, which involve higher powers of $h_{\mu\nu}$, describe the interaction of these perturbations.

Path integral formalism is a method used to quantise gravity models. It operates with the generation functional $\mathcal{Z}$, which is the following path integral:
\begin{align}
  \mathcal{Z} = \int \mathcal{D}[g] \,\exp\Big[ i\, \mathcal{A}[g_{\mu\nu}] \Big] .
\end{align}
The integration is performed over all possible spacetime metrics.

Gravity is a gauge theory, so it requires gauge fixing. The detailed discussion of the gauge fixing procedure lies far beyond the scope of this paper, and publications \cite{Latosh:2023zsi,Prinz:2020nru,Faddeev:1973zb} address it in full. This publication uses the gauge fixing procedure discussed in \cite{Latosh:2023zsi}. It influences the form of the graviton propagator but does not affect other results. In the next section, we briefly return to this discussion since the choice of gauge influences the form of the graviton propagator.

Perturbative expansion of the microscopic action yields the following expression for the generating functional:
\begin{align}
    \mathcal{Z} =& \int\mathcal{D}[\eta + \kappa \, h] \exp\left[ i\,\mathcal{A}[\eta_{\mu\nu}] + i\, \left. \frac{\delta\mathcal{A}}{\delta g_{\mu\nu}}\right|_{g=\eta} \, \kappa\, h_{\mu\nu} + i\, \left. \cfrac{\delta^2 \mathcal{A}}{ \delta g_{\mu\nu}\, \delta g_{\alpha\beta} } \right|_{g = \eta} \kappa^2\,h_{\mu\nu} h_{\alpha\beta} + \okappa{3}  \right]  \label{eq:Z1} \\
    =& \int\mathcal{D}[h] \exp\left[ \cfrac{i}{2} \, h^{\mu\nu} \mathcal{O}_{\mu\nu\alpha\beta} \square h^{\alpha\beta} + \kappa\, \left( \mathcal{V}^{(3)} \right)^{\mu_1\nu_1\mu_2\nu_2\mu_3\nu_3} h_{\mu_1\nu_1} h_{\mu_2\nu_2} h_{\mu_3\nu_3} + \okappa{2} \right]. \label{eq:Z2}
\end{align}
First and foremost, quantum degrees of freedom associated with $h_{\mu\nu}$ are referred to as gravitons. Second, the transition from the complete metric $\eta_{\mu\nu} + \kappa \, h_{\mu\nu}$ to the perturbations $h_{\mu\nu}$ in the integration measure has no influence on the path integral structure. The transition performs a finite shift of integration variables, so it does not influence the integral value. Third, the background term and the term linear in perturbation vanish. The background term is a constant and can be omitted. The term linear in perturbations vanishes since the $\eta_{\mu\nu}$ is a solution to the field equations generated by the microscopic action. Further, the term quadratic in perturbations describes the propagation of gravitons since it is quadratic in fields and derivatives. The operator $\mathcal{O}_{\mu\nu\alpha\beta}$ is known for the most part of gravity models \cite{Fierz:1939ix,VanNieuwenhuizen:1973fi,Accioly:2000nm,VanNieuwenhuizen:1981ae}. Lastly, all other terms involve three or more $h_{\mu\nu}$, so they describe interactions of gravitons. Because the microscopic action is an infinite series in perturbation, the corresponding quantum theory also contains an infinite number of interaction terms. However, all interaction terms share the same gravitational coupling $\kappa$. Within general relativity, a term describing the interaction of $n$ gravitons is multiplied on $\kappa^{n-2}$ since the action contains the Einstein-Hilbert part with the additional $\kappa^{-2}$ multiplier. The same may not hold for the other models involving more sophisticated couplings.

The microscopic action $\mathcal{A}$ generates differential operators describing the interaction of gravitons, such as operator $\left( \mathcal{V}^{(3)} \right)_{\mu_1\nu_1\mu_2\nu_2\mu_3\nu_3}$, which describes the interaction of three gravitons. Obtaining explicit expressions for these operators is challenging due to their complexity, often involving hundreds or thousands of terms \cite{DeWitt:1967yk,DeWitt:1967ub,DeWitt:1967uc,Prinz:2020nru,SevillanoMunoz:2022tfb,Latosh:2022ydd,Latosh:2023zsi}. We use the \texttt{FeynGrav} package \cite{Latosh:2022ydd,Latosh:2023zsi} to work with Feynman rules involving gravitons. \texttt{FeynGrav} is an extension of the widely used \texttt{FeynCalc} package \cite{Mertig:1990an,Shtabovenko:2016sxi,Shtabovenko:2020gxv}. We use \texttt{Package-X} and \texttt{FeynHelpers} for one-loop integral calculations \cite{Patel:2015tea,Patel:2016fam,Shtabovenko:2016whf,Passarino:1978jh}.

The perturbative approach to quantum gravity operates within the effective field theory paradigm. Effective field theory seeks to create a model applicable to the low-energy regime and does not aim to extend it to high energies. Perturbative gravity models are valid only for energies below the Planck scale because of the structure of the perturbation theory \cite{Burgess:2003jk,Donoghue:1994dn,Vanhove:2021zel}. The perturbative expansion \eqref{Microscopic_Action_Perturbative_Expansion} includes an infinite number of terms, all with the same gravitational coupling $\kappa$. The gravitational coupling is proportional to the inverse Planck scale $\kappa \sim m_\text{P}^{-1}$. Therefore, each term in the perturbative expansion \eqref{Microscopic_Action_Perturbative_Expansion} is suppressed by the energy factor $\left( E /  m_\text{P}\right)$, where $E$ is the typical interaction energy. The more interaction energy approaches the Planck scale, the more terms in the perturbative expansion become relevant, and the energy factor $\left( E /  m_\text{P}\right)$ approaches $1$. In the exact limit $E = m_\text{P}$, the perturbative expansion \eqref{Microscopic_Action_Perturbative_Expansion} diverges because perturbations become comparable to the background, so perturbation theory fails, and the model is no longer applicable.

The effective nature of perturbative quantum gravity allows the calculation of matrix elements despite its non-renormalizability. The theory is non-renormalisable because it introduces new operators at each level of perturbation theory. All ultraviolet-divergent factors can be subtracted at any given order of perturbation theory by introducing appropriate counter-terms. Each counter-term requires data to determine the value of its finite part. Thus, renormalising the theory at all orders of perturbation theory requires an infinite number of counter-terms and an infinite amount of data. However, renormalising an effective theory at all orders in perturbation theory is unnecessary. Regardless of the order of perturbation theory, the theory remains confined to the sub-Planckian domain and cannot be extended further. Consequently, one considers the first few orders of perturbation theory without the need for complete renormalisability.

The decoupling of scales is another crucial feature of an effective theory, clearly manifesting in perturbative quantum gravity. The decoupling of scales means that the model's low-energy behaviour is independent of its high-energy behaviour. Decoupling ensures that one can consistently study the theory at low energies, even without a complete ultraviolet theory of gravity. It manifests as follows. First, the theory enters the low-energy limit when all spatial momenta of external particles approach zero. In this case, the interaction energy also approaches zero, the energy factor $\left( E / m_\text{P} \right)$ remains small, and the theory stays perturbative. Second, the BPHZ theorem \cite{Bogoliubov:1957gp,Hepp:1966eg,Zimmermann:1969jj} defines the structure of all matrix elements in this limit. The theorem states that in any matrix element, the ultraviolet-divergent factor multiplies only an operator that is an analytic function of external momenta. Analytic functions are smooth and differentiable and remain bounded in the low-energy limit. Third, the theorem does not constrain the ultraviolet-finite terms, allowing them to be non-analytic (non-local) operators. Non-analytic operators can have various discontinuities and may become singular in the low-energy limit.

These factors lead to the following behaviour of matrix elements. One calculates a matrix element to a desirable order in perturbation theory and renormalises. After the renormalisation, all ultraviolet-divergent factors are replaced with finite constants that contain information about gravity's ultraviolet behaviour. These operators become subdominant in the low-energy limit because they are bounded and do not grow. In contrast, non-analytic operators grow uncontrollably and dominate low-energy behaviour. At the same time, these operators are unaffected by the renormalisation procedure and do not contain any information on the theory's high-energy behaviour. Consequently, in the low-energy limit, the leading contribution to any matrix element is ultraviolet finite and contains no information about high-energy gravitational physics.

The discussed feature of perturbative quantum gravity allows one to calculate the external gravitational field of a point-like particle. The calculation method dates back to \cite{Donoghue:1994dn}, and many publications discuss it in detail \cite{Donoghue:2001qc,Bjerrum-Bohr:2002fji,Bjerrum-Bohr:2002gqz,Bjerrum-Bohr:2014zsa,Bai:2016ivl,Chi:2019owc}. Due to the extensive discussion in literature, we only discuss its most essential steps.

The first major step is constructing a perturbative solution to the classical Einstein equations. One begins with the Einstein equations in the following form:
\begin{align}
  R_{\mu\nu} = 8\,\pi\,G_\text{N} \left[T_{\mu\nu} - \cfrac12\,g_{\mu\nu} \, T\,\right].
\end{align}
We aim to solve these equations for a point-like particle with the mass $M$. In order to construct a perturbative solution, one shall choose a small dimensionless parameter. In our case, there is a single suitable parameter $G_\text{N} M /r$ where $r$ is the distance from the particle. For the fixed value of mass $M$, this parameter is small when $r$ is large, so we construct an approximate solution describing the gravitational field far away from the particle.

One expands the metric, energy-momentum tensor, and all other quantities in a series with respect to small $G_\text{N} M/r$. The Einstein equations introduce an additional power of $G_\text{N}$ at the right-hand side, which mixes different orders of perturbation theory. In such a way, the equation relates the part of the Ricci tensor which is linear in perturbations ${}^{(1)} R_{\mu\nu}$ with the background part of the energy-momentum tensor ${}^{(0)} T_{\mu\nu}$ reads:
\begin{align}
  {}^{(1)}R_{\mu\nu} = 8\,\pi\,G_\text{N} \left[ {}^{(0)}T_{\mu\nu} - \cfrac12\,\eta_{\mu\nu}\,{}^{(0)}T \right]. 
\end{align}
The same equation takes the following form for the linear part of metric perturbations ${}^{(1)} g_{\mu\nu}$:
\begin{align}
    \square {}^{(1)} g_{\mu\nu} =- 8\,\pi\,G_\text{N}\, C_{\mu\nu\alpha\beta} \, {}^{(0)}T^{\alpha\beta},
\end{align}
where the $C_{\mu\nu\alpha\beta}$ is defined as follows:
\begin{align}
  C_{\mu\nu\alpha\beta} \overset{\text{def}}{=} \cfrac{1}{2} \bigg( \eta_{\mu\alpha}\eta_{\nu\beta} + \eta_{\mu\beta} \eta_{\nu\alpha} - \eta_{\mu\nu} \eta_{\alpha\beta} \bigg).
\end{align}
Since we consider the gravitational field of a point-like particle, we can always choose the frame where the particle rests, and the field is static. In that frame, all time dependence is removed, and one receives the Poisson equation:
\begin{align}
  \Delta {}^{(1)}g_{\mu\nu} \big( \vec{r} \, \big) = - 8\,\pi\,G_\text{N}\,C_{\mu\nu\alpha\beta}\, {}^{(0)}T^{\alpha\beta} \big( \vec{r} \, \big) .
\end{align}
The solution of this equation is well-known:
\begin{align}
  {}^{(1)}g_{\mu\nu}(\vec{r} \,)=& ~ 2\,G_\text{N} C_{\mu\nu\alpha\beta} \,\int d^3\vec{r'} ~ \cfrac{ {}^{(0)}T^{\alpha\beta}(\vec{r'}) }{\abs{\vec{r}-\vec{r'}}} \, .
\end{align}
For our purposes, it is more convenient to perform the Fourier transformation and operate with the momentum representation of the energy-momentum tensor:
\begin{align}
    {}^{(1)}g_{\mu\nu}(\vec{r}\,) & = 2\,G_\text{N}\,C_{\mu\nu\alpha\beta}\, \int \cfrac{d^3\vec{k}}{(2\pi)^3} \, e^{-i\vec{k}\cdot\vec{r}} \, \cfrac{1}{\abs{\vec{k}}^2} \, {}^{(0)}T^{\alpha\beta} \Big( \vec{k} \Big), \label{Linear_Part_of_Perturbative_Metric} \\
    {}^{(0)}T^{\alpha\beta} \Big( \vec{k} \Big) & \overset{\text{def}}{=} \int d^3 \vec{r} ~ {}^{(0)}T^{\alpha\beta}(\vec{r} \,) e^{i\,\vec{k} \cdot \vec{r}}.
\end{align}

The second step is to discover the relation between this solution and the low-energy limit to find a way to relate classical and quantum descriptions. The formula \eqref{Linear_Part_of_Perturbative_Metric} describes the leading part of the gravitational field detected by a distant observer far removed from the particle in the classical theory. At the same time, the observer weakly interacts with the particle and does not experience a relative motion. Consequently, the interaction energy approaches zero while the spacial momenta of the particle and observer and the particle vanish, so the system enters the low-energy regime described above. Since the classical theory enters the validity region of perturbative quantum gravity, one can generalise \eqref{Linear_Part_of_Perturbative_Metric}.

We use the following formula to describe the leading order contribution contribution to the gravitational field of a massive quantum particle:
\begin{align}\label{the_connection_formula}
  {}^{(1)}g_{\mu\nu}(\vec{r}\,)=2\,G_\text{N}\,C_{\mu\nu\alpha\beta}\, \int \cfrac{d^3 \vec{k}}{(2\pi)^3} \, e^{-i\vec{k}\cdot\vec{r}} \, \cfrac{1}{\abs{\vec{k}}^2} \, \left\langle \widehat{T}^{\alpha\beta}(\vec{k}) \right\rangle .
\end{align}
In this equation, $\left\langle \widehat{T}^{\alpha\beta}(\vec{k}) \right\rangle$ is a matrix element describing the observed value of the energy-momentum tensor. 

The definition of the energy-momentum tensor used in \eqref{the_connection_formula} is more subtle than in the classical case. In the classical case, the following formula gives the energy-momentum tensor of matter described by Lagrangian $\mathcal{L}_\text{matter}$:
\begin{align}
    T_{\mu\nu}^\text{classical} \overset{\text{def}}{=}  \cfrac{(-2)}{\sqrt{-g}} \, \cfrac{\delta}{\delta g^{\mu\nu}} \left( \sqrt{-g} \, \mathcal{L}_\text{matter} \right).
\end{align}
This formula produces expressions that contain both $g^{\mu\nu}$ and $\sqrt{-g}$. For the perturbative metric \eqref{perturbative_metric}, these factors are infinite series in $\kappa$, so the classical expression for the energy-momentum tensor is an infinite series which terms contain powers of $h_{\mu\nu}$. The perturbation $h_{\mu\nu}$ corresponds to the graviton operator in perturbative quantum gravity. Hence, directly promoting the classical energy-momentum tensor to the operator produces an infinite series in graviton operators. This expression cannot be consistently interpreted since it mixes matter and graviton states.

Within perturbative quantum gravity, one shall examine the structure of the matter coupling to gravity in more detail. The microscopic action describing matter involves $g^{\mu\nu}$ and $\sqrt{-g}$, producing an infinite series of perturbations. We can use the following formula to describe its structure:
\begin{align}
    \int\!\! d^4 x \sqrt{-g} \mathcal{L}_\text{matter} = \!\! \int\!\! d^4 x \Bigg[ \mathcal{L}^{(0)}_\text{matter}\!\! + \! \kappa T^{\mu\nu} h_{\mu\nu} \! + \! \kappa^2 T^{\mu\nu\alpha\beta} h_{\mu\nu} h_{\alpha\beta} \! + \! \cdots \! + \! \kappa^n T^{\rho_1\sigma_1\cdots\rho_n\sigma_n} h_{\rho_1\sigma_1} \! \cdots \! h_{\rho_n\sigma_n} \! +\! \cdots \Bigg].
\end{align}
In this expression, $\mathcal{L}_\text{matter}$ is the full Lagrangian describing matter that involves both $g_{\mu\nu}$ and  $g^{\mu\nu}$. The background contribution $\mathcal{L}^{(0)}_\text{matter}$ evaluated at the flat metric $\eta_{\mu\nu}$, so it does not contain $h_{\mu\nu}$:
\begin{align}
    \mathcal{L}^{(0)}_\text{matter} \overset{\text{def}}{=} \, \sqrt{-g}\, \mathcal{L}_\text{matter} \Bigg|_{g_{\mu\nu} = \eta_{\mu\nu}}
\end{align}
This background contribution describes the matter dynamics without gravity. It describes the propagation of matter states and their interactions. The next term describes a linear coupling of matter to gravity, so the tensor $T^{\mu\nu}$ defined via a variational derivative
\begin{align}
    T^{\mu\nu} \overset{\text{def}}{=} \cfrac{\delta}{\delta g_{\mu\nu}} \left( \, \sqrt{-g}\, \mathcal{L}_\text{matter} \right) \Bigg|_{g_{\mu\nu} = \eta_{\mu\nu}}.
\end{align}
The corresponding term in the action describes a coupling of matter degrees of freedom to a single graviton. The next term describes the quadratic coupling to small metric perturbations, so the tensor $T^{\mu\nu\alpha\beta}$ is a second variational derivative evaluated at the flat background:
\begin{align}
    T^{\mu\nu\alpha\beta} \overset{\text{def}}{=} \cfrac{\delta}{\delta g_{\mu\nu}} \, \cfrac{\delta}{\delta g_{\alpha\beta}} \,\left( \, \sqrt{-g}\, \mathcal{L}_\text{matter} \right) \Bigg|_{g_{\mu\nu} = \eta_{\mu\nu}}.
\end{align}
In turn, the corresponding part of the action describes the interaction with two gravitons. The same logic applies to all subsequent terms, so the tensor
\begin{align}
    T^{\rho_1\sigma_1\cdots\rho_n\sigma_n} \overset{\text{def}}{=} \cfrac{\delta}{\delta g^{\rho_1\sigma_1}}\cdots\cfrac{\delta}{\delta g^{\rho_n\sigma_n}} \, \left( \, \sqrt{-g}\, \mathcal{L}_\text{matter} \right) \Bigg|_{g_{\mu\nu} = \eta_{\mu\nu}}
\end{align}
describes a coupling of matter to $n$ gravitons. These operators are calculated within perturbative quantum gravity for many models \cite{Latosh:2022ydd,Latosh:2023zsi}, and they enter the corresponding Feynman rules. However, formula \eqref{the_connection_formula} uses a different notion of the energy-momentum tensor.

To define the operator $\widehat{T}^{\alpha\beta}$, we shall first recall that we address the gravitational field of a single particle. To calculate the gravitational field, one shall consider a process when a one-particle on-shall state radiates an off-shell graviton \cite{Donoghue:1994dn}. In that sense, we consider a coupling of matter states to the gravitational field, but our case is principally different from both cases discussed above. In contrast with the classical case, we consider a coupling to a single graviton $h_{\mu\nu}$, not to the $g^{\mu\nu}$ or $\sqrt{-g}$. In contrast with the $T^{\rho_1\sigma_1\cdots\rho_n\sigma_n}$ tensors that enter the Feynman rules, we describe a coupling to a single graviton at all orders of perturbation theory. 

The given reasoning shows that the operator $\widehat{T}^{\alpha\beta}$ accounts for all processes describing a transition of a $3$-momentum $\vec{k}$ from a one-particle on-shell state to an off-shell graviton. The particle must be on the mass shell since it is a real resting particle. The graviton must be placed off the mass shell to describe the interaction. The matrix element would describe gravitational radiation if the graviton were placed on the mass shell. The right-hand side of \eqref{the_connection_formula} shall be calculated within the low-energy limit when the transferred momentum approaches zero. The decoupling of scales ensures that the result of such calculations does not depend on the high-energy behaviour of the model.

It is useful to show the calculation of $\langle \widehat{T}_{\mu\nu}(\vec{k})\rangle$ within general relativity \cite{Donoghue:1994dn}. In full accordance with the given logic, one shall account for all processes contributing to the matrix element at different orders of the perturbation theory:
\begin{align}\label{Series_1}
\langle \widehat{T}_{\mu\nu}(\vec{k})\rangle = \hspace{20pt}
  \begin{gathered}
    \begin{fmffile}{Series_1_D1}
      \begin{fmfgraph*}(30,30)
        \fmfleft{L}
        \fmfright{R1,R2}
        \fmf{dbl_wiggly}{L,V}
        \fmf{dashes}{R1,V,R2}
        \fmfv{decoration.shape=circle,decoration.size=15pt,decoration.filled=shaded}{V}
        \fmflabel{$\mu\nu$}{L}
      \end{fmfgraph*}
    \end{fmffile}
  \end{gathered}
  =
  \begin{gathered}
    \begin{fmffile}{Series_1_D2}
      \begin{fmfgraph}(30,30)
        \fmfleft{L}
        \fmfright{R1,R2}
        \fmf{dbl_wiggly}{L,V}
        \fmf{dashes}{R1,V,R2}
        \fmfdot{V}
      \end{fmfgraph}
    \end{fmffile}
  \end{gathered}
  +
  \begin{gathered}
    \begin{fmffile}{Series_1_D3}
      \begin{fmfgraph}(30,30)
        \fmfleft{L}
        \fmfright{R1,R2}
        \fmf{dbl_wiggly,tension=2}{L,V}
        \fmf{dashes,tension=2}{R1,V1}
        \fmf{dashes,tension=2}{R2,V2}
        \fmf{dashes}{V1,V}
        \fmf{dashes}{V2,V}
        \fmffreeze
        \fmf{dbl_wiggly}{V1,V2}
        \fmfdot{V,V1,V2}
      \end{fmfgraph}
    \end{fmffile}
  \end{gathered}
  +
  \begin{gathered}
    \begin{fmffile}{Series_1_D4}
      \begin{fmfgraph}(30,30)
        \fmfleft{L}
        \fmfright{R1,R2}
        \fmf{dbl_wiggly,tension=2}{L,V}
        \fmf{dashes,tension=2}{R1,V1}
        \fmf{dashes,tension=2}{R2,V2}
        \fmf{dbl_wiggly}{V1,V}
        \fmf{dbl_wiggly}{V2,V}
        \fmffreeze
        \fmf{dashes}{V1,V2}
        \fmfdot{V,V1,V2}
      \end{fmfgraph}
    \end{fmffile}
  \end{gathered}
  +
  \begin{gathered}
    \begin{fmffile}{Series_1_D5}
      \begin{fmfgraph}(30,30)
        \fmfleft{L}
        \fmfright{R1,R2}
        \fmf{dashes,tension=2}{R1,VR,R2}
        \fmf{phantom,tension=0.7}{L,VR}
        \fmffreeze
        \fmf{dbl_wiggly}{L,V}
        \fmf{phantom}{V,VR}
        \fmffreeze
        \fmf{dbl_wiggly,right=1}{V,VR,V}
        \fmfdot{VR,V}
      \end{fmfgraph}
    \end{fmffile}
  \end{gathered}
  +
  \begin{gathered}
    \begin{fmffile}{Series_1_D6}
      \begin{fmfgraph}(30,30)
        \fmfleft{L}
        \fmfright{R1,R2}
        \fmf{dbl_wiggly,tension=2}{L,V}
        \fmf{dashes,tension=0.5}{R1,V}
        \fmf{dashes,tension=0.5}{R2,V}
        \fmffreeze
        \fmf{phantom,tension=2}{R1,V1}
        \fmf{phantom}{V1,V}
        \fmffreeze
        \fmf{dbl_wiggly,left=0.7}{V,V1}
        \fmfdot{V,V1}
      \end{fmfgraph}
    \end{fmffile}
  \end{gathered}
  +
  \begin{gathered}
    \begin{fmffile}{Series_1_D7}
      \begin{fmfgraph}(30,30)
        \fmfleft{L}
        \fmfright{R1,R2}
        \fmf{dbl_wiggly,tension=2}{L,V}
        \fmf{dashes,tension=0.5}{R1,V}
        \fmf{dashes,tension=0.5}{R2,V}
        \fmffreeze
        \fmf{phantom,tension=2}{R2,V1}
        \fmf{phantom}{V1,V}
        \fmffreeze
        \fmf{dbl_wiggly,right=0.7}{V,V1}
        \fmfdot{V,V1}
      \end{fmfgraph}
    \end{fmffile}
  \end{gathered}
  +
  \begin{gathered}
    \begin{fmffile}{Series_1_D8}
      \begin{fmfgraph}(30,30)
        \fmfleft{L}
        \fmfright{R1,R,R2}
        \fmf{dbl_wiggly,tension=2}{L,V}
        \fmf{dashes,tension=0.5,left=0.5}{R1,V,R2}
        \fmffreeze
        \fmf{phantom,tension=3}{R,VR}
        \fmf{phantom}{V,VR}
        \fmffreeze
        \fmf{dbl_wiggly,right=0.7}{V,VR,V}
        \fmfdot{V}
      \end{fmfgraph}
    \end{fmffile}
  \end{gathered}
  +\okappa{5}.
\end{align}
On the left-hand side of this expression is a diagram denoting the complete matrix element that accounts for all orders of perturbation theory. The dashed lines in this and other diagrams correspond to massive scalar particles, while the double wiggly lines correspond to gravitons. On the right hand, the first diagram corresponds to the leading order contribution (tree-level), while the others describe next-to-leading order corrections (one-loop level). In full agreement with the discussion given above, one shall calculate these corrections and take the low-energy limit, making the spatial momenta of the scalar vanish.

The procedure relating the matrix element with the metric describing the external field of a particle was developed in \cite{Donoghue:1994dn,Donoghue:2001qc}, where it is discussed in detail. From the practical point of view, the derivation only introduces form factors that help one to obtain an expression for the metric with the equation \eqref{the_connection_formula}. One introduces form factors $F_1$ and $F_2$ that describe the universal structure of the matrix element
\begin{align}
    \langle \widehat{T}_{\mu\nu}(\vec{k})\rangle = i\, \cfrac{\kappa}{2} F_1(k^2) \left[ (p_1)_\mu (p_2)_\nu + (p_1)_\nu (p_2)_\mu - \cfrac12\, k^2 \, \eta_{\mu\nu} \right] + i\, \cfrac{\kappa}{2} F_2(k^2) \left( k_\mu k_\nu - \eta_{\mu\nu} k^2\right) ,
\end{align}
After this, equation \eqref{the_connection_formula} gives the following explicit expression for metric components:
\begin{align}\label{Donoghue_metric_formula}
    \begin{cases}
        {}^{(1)}g_{00} &= -16\,\pi\, G_\text{N} \int \, \cfrac{d^3 \vec{k}}{(2\pi)^3} e^{i \vec{k}\cdot\vec{r} } \cfrac{1}{\abs{\vec{k}}} \left( \cfrac{m}{2} F_1\left(-\vec{k}^2\right) - \cfrac{\vec{k}^2}{4\,m} \, F_2\left(-\vec{k}^2\right)  \right) \,, \\
        {}^{(1)}g_{0i} &= 0 \,, \\
        {}^{(1)}g_{ij} &= -16\,\pi\, G_\text{N} \int \, \cfrac{d^3 \vec{k}}{(2\pi)^3} e^{i \vec{k}\cdot\vec{r} } \cfrac{1}{\abs{\vec{k}}} \left( \cfrac{m}{2} F_1\left(-\vec{k}^2\right) + \cfrac{1}{2\,m} \left( k_i k_j + \cfrac12\,\delta_{ij} \abs{\vec{k}}^2\right) F_2\left(-\vec{k}^2\right) \right) \,. \\
    \end{cases}
\end{align}

One obtains the two-body potential similarly. Similarly to the previous case, the calculation method was developed in \cite{Donoghue:1994dn,Donoghue:2001qc}, and we only recall its essential steps. First, one can extract a two-body interaction potential from the $2 \to 2$ scattering matrix element:
\begin{align}\label{{Donoghue_potential_formula}}
\mathcal{M} = 
    \begin{gathered}
        \begin{fmffile}{Series_6_D1}
            \begin{fmfgraph}(40,40)
            \fmfleft{L1,L2}
            \fmfright{R1,R2}
            \fmf{dashes}{L1,VL}
            \fmf{dashes}{L2,VL}
            \fmf{dashes}{R1,VR}
            \fmf{dashes}{R2,VR}
            \fmf{dbl_wiggly,tension=0.7}{VL,VR}
            \fmfv{decoration.shape=circle,decoration.size=10pt,decoration.filled=shaded}{VL,VR}
            \end{fmfgraph}
        \end{fmffile}
    \end{gathered}
\end{align}
In this expression, one accounts for all vertex corrections \eqref{Series_1}. At the tree level, the matrix element matches the expression for the Newtonian two-body potential:
\begin{align}
    \begin{split}
        \mathcal{M} \left( \vec{k} \right) &= - \cfrac{\kappa^2}{8} \, \cfrac{m_1\,m_2}{ \abs{\vec{k}}^2 } + \okappa{4} , \\
        \mathcal{M} \left( \vec{r} \right) &= \int \cfrac{d^3 \vec{k}}{(2\pi)^3} \, e^{i\,\vec{k} \cdot \vec{r}} ~ \mathcal{M} \left( \vec{k} \right)  = - \cfrac{G_\text{N} \, m_1\, m_2}{r} + \mathcal{O} \left(G_\text{N}^2 \right) .
    \end{split}
\end{align}
In the case of general relativity, the next-to-leading order corrections read \cite{Donoghue:1994dn}:
\begin{align}
    \mathcal{M} \left( \vec{r} \right) = \cfrac{G_\text{N} \, m_1 \, m_2}{r} \left[ 1 - \cfrac{G_\text{N} (m_1 + m_2) }{r} - \cfrac{127 \hbar}{30\,\pi^2} \,\cfrac{G_\text{N}}{r^2 c^3} + \mathcal{O}\left(G_\text{N}\right)^2 \right] .
\end{align}

These results suggest that, for a distant observer, the gravitational field of a point-like particle takes the same form as the gravitational field of a black hole in a few leading orders in perturbative expansion. Both a particle and a black hole are characterised by their mass, spin, and charge. Nevertheless, this does not imply an association of particles with black holes. Instead, it demonstrates that the gravitational field of a particle serves as a suitable model for the gravitational field of a macroscopic body at large distances, where the body's internal structure can be neglected.

These calculation methods apply similarly to other gravity models. The only difference is the new interactions present in the theory. When the theory introduces new interactions, one must revisit the perturbation theory, which produces different diagrams in next-to-leading order and beyond. The following section discusses the implementation of this method to the discussed models with the non-minimal coupling to the Gauss-Bonnet term.

This section summarises as follows. First, the perturbative quantum gravity provides a way to study quantum gravitational effects consistently. The theory operates within the effective field theory paradigm, so it is applicable below the Planck scale. The decoupling of scale allows one to study features of the theory that are independent of its possible ultraviolet completion. Secondly, the theory provides a way to study the gravitational field of a particle \cite{Donoghue:1994dn,Donoghue:2001qc}. One can calculate the two-body interaction potential and the metric describing the gravitational field of a single particle to any order in perturbation theory. Thirdly, within general relativity, such calculations recover the perturbative expansion of the Schwarzshild metric for a massive scalar particle, the Kerr metric for the massive fermion, and the Kerr-Newman metric for the massive charged fermion. This methodology is employed throughout this paper. We compute the leading corrections to the gravitational field of a particle induced by the non-minimal coupling to the Gauss-Bonnet term. We apply these corrections to describe the impact of non-minimal coupling on the gravitational field of a macroscopic body.

\section{External gravitational field}\label{Section_metric_reconstruction}

This section applies the perturbative quantum gravity formalism to scalar-tensor gravity models with non-minimal coupling to the Gauss-Bonnet term. Firstly, we slightly change the model \eqref{the_action} to make it more suitable for the perturbative quantum gravity framework. Second, we discuss the Feynman rules of the model and construct the amplitudes to recover the external gravitational field of a particle. Lastly, we calculate the leading corrections to the non-relativistic potential and the metric generated by the non-minimal coupling.

We begin with the discussion of the model \eqref{the_action}. First and foremost, we shall introduce the mass term for the scalar field. Secondly, we shall introduce a different parametrisation of the non-minimal coupling to the Gauss-Bonnet term. We discuss the reason for choosing the new parameterisation in the following paragraph. From here on, we consider a model given by the following action:
\begin{align}\label{the_action_2}
  S = \int d^4 x \sqrt{-g} \left[ -\cfrac{2}{\kappa^2}\,R + \cfrac12\,g^{\mu\nu}\, \nabla_\mu\varphi\, \nabla_\nu \varphi  - \frac{1}{2} m^2 \varphi^2 - \cfrac{1}{\kappa^2}\,\cfrac{1}{\mathcal{E}_{\varphi}^2} \, F\left( \cfrac{\varphi}{\mathcal{E}_{\varphi}} \right) \,\mathcal{G}  \right] .
\end{align}
In this expression, $m$ is a scalar field mass, and $\mathcal{E}_{\varphi}$ is a new energy scale of the non-minimal coupling.

We use \eqref{the_action_2} because it introduces a single new energy scale, while the original action \eqref{the_action} introduces several independent energy scales. The quantum field theory formalism describes interactions of particles, and it can only operate with expressions involving whole powers of fields. To promote action \eqref{the_action} to quantum field theory, one shall expand $f$ in the Taylor series with respect to small $\varphi$, so each term of this series describes an interaction with a given number of scalars. Namely, if the expansion for the function $f$ reads
\begin{align}
    f(\varphi) = f_0 + f_1 \varphi + \cfrac{1}{2!}\,f_2 \varphi^2 + \cdots,
\end{align}
then $f_1$ describes the non-minimal interaction of one scalar with gravity, $f_2$ describes the non-minimal interaction of two scalars, and so on.

This parametrisation causes the problem because each coefficient has a dimension and defines a new energy scale. Those energy scales are independent since one does not impose any conditions on $f$. To put it otherwise, the model \eqref{the_action} describes not a single non-minimal coupling between a scalar and the Gauss-Bonnet term but a series of independent non-minimal couplings. The presence of many independent energy scales makes controllable calculations impossible. Since energy interaction scales are independent, controlling which interactions provide the leading contribution and which are neglectable is impossible.

Action \eqref{the_action_2} avoids this issue entirely because it fixes a single interaction scale. First, the dimensionless multiplier $(\kappa \, \mathcal{E}_\varphi)^{-2}$ ensures that the function $F$ is dimensionless. Secondly, the function $F$ explicitly depends on a dimensionless argument $\varphi/ m_\text{Pl}$. Because of these two factors, the function $F$ defines and enforces a single energy scale on each term in the Taylor series.

Action \eqref{the_action_2} also uniquely solves the interaction hierarchy problem. Only two hierarchy cases exist since the models introduce a single new interaction scale. The first case is when the non-minimal coupling is suppressed more strongly than the standard gravitational interaction, i.e. $\mathcal{E}_{\varphi} > m_\text{Pl}$. This case is irrelevant because it lies beyond the scope of effective field theory applicability. The effective theory is applicable only below the Planck scale. If the new interaction is suppressed even strongly, it can safely be neglected for the whole sup-Planckian region, where we can use the effective theory. The second case is when the non-minimal coupling experiences a weaker suppression than the standard gravitational interaction. In that case, the non-minimal coupling becomes relevant below the Planck scale and shall be accounted for within the effective field theory applicability region. The standard gravitational interaction is strongly suppressed and can only provide the leading order contribution. This case is physically relevant, and we will consider it further.

Further calculations follow the perturbative quantum gravity scheme described above. One defines the complete metric \eqref{perturbative_metric} in terms of small perturbations and expands the microscopic action \eqref{the_action_2} in an infinite perturbative series. To obtain the graviton propagator, we follow \cite{Latosh:2022ydd} and introduce the de Donder gauge fixing term:
\begin{align}
    \mathcal{A}_\text{gf} = \cfrac{1}{\kappa^2} \, \int d^4 x \sqrt{-g} \,  g_{\mu\nu} \, \Big( g^{\alpha\beta} \Gamma^\mu_{\alpha\beta} \Big) \Big( g^{\rho\sigma} \Gamma^\nu_{\rho\sigma} \Big) .
\end{align}
It results in the following graviton propagator:
\begin{align}\label{Graviton_Propagator}
    \begin{gathered}
        \begin{fmffile}{Graviton_Propagator}
            \begin{fmfgraph*}(30,30)
                \fmfleft{L}
                \fmfright{R}
                \fmf{dbl_wiggly}{L,R}
                \fmflabel{$\mu\nu$}{L}
                \fmflabel{$\alpha\beta$}{R}
            \end{fmfgraph*}
        \end{fmffile}
    \end{gathered} \hspace{20pt} = \cfrac{i}{2} \, \cfrac{ \eta_{\mu\alpha} \eta_{\nu\beta} + \eta_{\mu\beta}\eta_{\nu\alpha} - \eta_{\mu\nu} \eta_{\alpha\beta} }{k^2} \, .
\end{align}
The Faddeev-Popov ghosts do not contribute to the diagrams discussed below. They will contribute to a higher level of perturbation theory, so we do not discuss them further.

The Feynman rules describing the minimal coupling of a scalar field to gravity are discussed in many publications \cite{Latosh:2021usy,Latosh:2022ydd,Latosh:2023zsi}, so we present them without a derivation:
\begin{align}\label{Graviton_Rule_1}
    \nonumber \\
    \begin{gathered}
        \begin{fmffile}{Graviton_Rule_1}
            \begin{fmfgraph*}(30,30)
                \fmfleft{L}
                \fmfright{R1,R2}
                \fmf{dashes}{R1,V}
                \fmf{dashes}{R2,V}
                \fmf{dbl_wiggly}{V,L}
                \fmfdot{V}
                \fmflabel{$\mu\nu$}{L}
                \fmflabel{$p_1$}{R1}
                \fmflabel{$p_2$}{R2}
            \end{fmfgraph*}
        \end{fmffile}
    \end{gathered}
    & = \cfrac{i}{2} \,\kappa\left[ (p_1)_\mu (p_2)_\nu + (p_1)_\nu (p_2)_\mu - \eta_{\mu\nu} (p_1\cdot p_2) - m^2 \, \eta_{\mu\nu}  \right] ;\\ \nonumber \\ \nonumber \\
    \begin{gathered}\label{Graviton_Rule_2}
        \begin{fmffile}{Graviton_Rule_2}
            \begin{fmfgraph*}(30,30)
                \fmfleft{L1,L2}
                \fmfright{R1,R2}
                \fmf{dashes}{R1,V}
                \fmf{dashes}{R2,V}
                \fmf{dbl_wiggly}{V,L1}
                \fmf{dbl_wiggly}{V,L2}
                \fmfdot{V}
                \fmflabel{$\mu\nu$}{L1}
                \fmflabel{$\alpha\beta$}{L2}
                \fmflabel{$p_1$}{R1}
                \fmflabel{$p_2$}{R2}
            \end{fmfgraph*}
        \end{fmffile}
    \end{gathered}
    & = \cfrac{i}{8}\,\kappa^2 \Bigg[ \left( m^2 + p_1 \cdot p_2 \right) \left( \eta_{\alpha\nu} \eta_{\beta\mu} + \eta_{\alpha\mu} \eta_{\beta\nu} - \eta_{\alpha\beta} \eta_{\mu\nu} \right)  \\
    & \hspace{35pt} + \left( (p_1)_\beta (p_1)_\alpha + (p_1)_\alpha (p_1)_\beta \right) \eta_{\mu\nu} + \left( (p_1)_\nu (p_1)_\mu + (p_1)_\mu (p_1)_\nu \right) \eta_{\alpha\beta} \nonumber \\
    & \hspace{35pt}- \left( (p_1)_\nu (p_1)_\alpha + (p_1)_\alpha (p_1)_\nu \right) \eta_{\beta\mu} - \left( (p_1)_\mu (p_1)_\alpha + (p_1)_\alpha (p_1)_\mu \right) \eta_{\beta\nu} \nonumber\\
    & \hspace{35pt}- \left( (p_1)_\mu (p_1)_\beta + (p_1)_\beta (p_1)_\mu \right) \eta_{\alpha\nu} - \left( (p_1)_\nu (p_1)_\beta + (p_1)_\beta (p_1)_\nu \right) \eta_{\alpha\mu} ~ \Bigg]. \nonumber
\end{align}

The interaction rules for the non-minimal coupling to the Gauss-Bonnet term are derived as follows. We expand the Gauss-Bonnet coupling to the leading order in metric perturbations:
\begin{align}
  \int d^4 x \sqrt{-g} \, \cfrac{1}{\kappa^2}\,\cfrac{1}{\mathcal{E}^2}\, F\left(\cfrac{\varphi}{\mathcal{E}}\right) \,\mathcal{G} =\cfrac{1}{\mathcal{E}^2}\, \int d^4 x \,F\left( \cfrac{\varphi}{\mathcal{E}} \right) \, \widehat{\mathcal{G}}^{\mu\nu\alpha\beta} h_{\mu\nu} h_{\alpha\beta} +\okappa{1}.
\end{align}
In this expression, we defined a new differential operator $\widehat{\mathcal{G}}^{\mu\nu\alpha\beta}$. The derivation of this operator involves only technical computations, and we place the explicit expression for the operator \eqref{Gauss-Bonnet_tensor_structure} together with its derivation in Appendix \ref{Appendix_Gauss-Bonnet}.

As noted above, to construct a quantum theory for model \eqref{the_action_2}, one must expand the function $F$ in the Taylor series:
\begin{align}
  F \left(\cfrac{\varphi}{\mathcal{E}} \right) = \sum\limits_{n=1}^{\infty} \cfrac{F_n}{n!} \, \left( \cfrac{\varphi}{\mathcal{E}} \right)^n \,.
\end{align}
Each term in this series is responsible for an interaction of $n$ scalar particles coupled to the Gauss-Bonnet term. For practical purposes, we can consider a single term of this series since the interaction rules for each term are obtained similarly. Specifically, we have
\begin{align}
  \left. \cfrac{1}{\kappa^2}\,\cfrac{1}{\mathcal{E}^2}\, \int d^4 x \sqrt{-g} \,\cfrac{F_n}{n!} \, \left(\cfrac{\varphi}{\mathcal{E}}\right)^n \,\mathcal{G} \right|_\text{Leading Order} = \cfrac{1}{\mathcal{E}^{n+2}}\,\cfrac{F_n}{n!} \int d^4 x\, ~ \varphi^n ~ \widehat{\mathcal{G}}^{\mu\nu\alpha\beta} \,h_{\mu\nu} h_{\alpha\beta}.
\end{align}
This results in the following interaction rule, which is marked with a hollow circle to distinguish it from the pure general relativity interaction vertices:
\begin{align}\label{Graviton_Rule_3}
  \nonumber \\
  \begin{gathered}
    \begin{fmffile}{Vertex_GB}
      \begin{fmfgraph*}(40,40)
        \fmfleft{L1,L2}
        \fmfright{R1,R2}
        \fmf{dashes}{L1,V}
        \fmf{dashes}{L2,V}
        \fmf{dbl_wiggly}{V,R1}
        \fmf{dbl_wiggly}{V,R2}
        \fmffreeze
        \fmf{dots}{L1,L2}
        \fmflabel{$p_1$}{L1}
        \fmflabel{$p_n$}{L2}
        \fmflabel{$\mu\nu,k_1$}{R1}
        \fmflabel{$\alpha\beta,k_2$}{R2}
        \fmfv{decoration.shape=circle,decoration.filled=empty,decoration.size=7}{V}
      \end{fmfgraph*}
    \end{fmffile}
  \end{gathered}
  \hspace{20pt}=i ~ \cfrac{1}{\mathcal{E}^{n+2}} ~ F_n \,\widehat{\mathcal{G}}^{\mu\nu\alpha\beta} (k_1, k_2). \\ \nonumber
\end{align}
This expression is symmetric with respect to graviton indices and by permutation of both gravitons and scalars.

Formulae \eqref{Graviton_Propagator}, \eqref{Graviton_Rule_1}, \eqref{Graviton_Rule_2}, and \eqref{Graviton_Rule_3} give us the complete set of Feynman rules for the model. We shall specify the coupling function $F$ and use the interaction hierarchy discussed above to proceed with the calculations. 

We begin with the linear coupling case. In terms of \eqref{the_action} parametrisation the coupling function reads:
\begin{align}\label{the_linear_coupling_definition}
  f(\varphi) = \lambda\, \varphi .
\end{align}
Here $\lambda$ is a coupling with the mass dimension $-1$. In terms of parametrisation \eqref{the_action_2}, the expression describing the coupling to the Gauss-Bonnet term becomes
\begin{align*}
    \int d^4 x \sqrt{-g} \Bigg[ - \cfrac{1}{\kappa^2 \, \mathcal{E}^2} ~ \cfrac{\varphi}{\mathcal{E}} ~ \mathcal{G} \Bigg]\, .
\end{align*}
The interaction scale $\mathcal{E}$ is related with the coupling $\lambda$:
\begin{align}
  \mathcal{E} = \cfrac{1}{ \sqrt[3]{\lambda\, \kappa^2}} \,.
\end{align}

In full accordance with the hierarchy reasoning given above, the interaction energy scale $\mathcal{E}$ shall be below the Planck scale for the interaction to be correctly described by the perturbation theory. In turn, the standard gravitational interaction experiences a stronger suppression, and its higher-order correction is neglected. Consequently, only the following diagrams contribute to the matrix element responsible for the low-energy limit:
\begin{align}
  \nonumber \\
  \begin{gathered}
    \begin{fmffile}{Series_2_D1}
      \begin{fmfgraph*}(30,30)
        \fmfleft{L}
        \fmfright{R1,R2}
        \fmf{dbl_wiggly}{L,V}
        \fmf{dashes}{R1,V,R2}
        \fmfv{decoration.shape=circle,decoration.size=15pt,decoration.filled=shaded}{V}
        \fmflabel{$\mu\nu,k$}{L}
        \fmflabel{$p_1$}{R1}
        \fmflabel{$p_2$}{R2}
      \end{fmfgraph*}
    \end{fmffile}
  \end{gathered}
  \hspace{10pt} = ~\Bigg[ ~~
  \begin{gathered}
    \begin{fmffile}{Series_2_D2}
      \begin{fmfgraph}(30,30)
        \fmfleft{L}
        \fmfright{R1,R2}
        \fmf{dbl_wiggly}{L,V}
        \fmf{dashes}{R1,V,R2}
        \fmfdot{V}
      \end{fmfgraph}
    \end{fmffile}
  \end{gathered}
  + ~
  \begin{gathered}
    \begin{fmffile}{Series_2_D3}
      \begin{fmfgraph}(30,30)
        \fmfleft{L}
        \fmfright{R1,R2}
        \fmf{dbl_wiggly,tension=2}{L,VL}
        \fmf{dashes,tension=2}{R1,V1}
        \fmf{dashes,tension=2}{R2,V2}
        \fmf{dbl_wiggly,tension=0.5}{VL,V1,V2,VL}
        \fmfdot{VL}
        \fmfv{decoration.shape=circle,decoration.filled=empty,decoration.size=5}{V1,V2}
      \end{fmfgraph}
    \end{fmffile}
  \end{gathered}
  +\mathcal{O}\left( \mathcal{E}^{-3} \right) \Bigg] + \okappa{3} . \\ \nonumber
\end{align}
In this diagram, $p_1$ and $p_2$ are momenta of on-shell scalars, and $k$ is the graviton momentum. The first diagram describes the standard gravitational coupling that produces Newtonian potential. The second diagram generates the power law corrections in $r$ to the Newtonian potential.

Let us discuss the structure of the loop diagram:
\begin{align}
  \nonumber \\
  \begin{gathered}
    \begin{fmffile}{D1}
      \begin{fmfgraph*}(30,30)
        \fmfleft{L}
        \fmfright{R1,R2}
        \fmf{dbl_wiggly,tension=2}{L,VL}
        \fmf{dashes,tension=2}{R1,V1}
        \fmf{dashes,tension=2}{R2,V2}
        \fmf{dbl_wiggly,tension=0.5}{VL,V1,V2,VL}
        \fmfdot{VL}
        \fmfv{decoration.shape=circle,decoration.filled=empty,decoration.size=5}{V1,V2}
        \fmflabel{$\mu\nu,k$}{L}
        \fmflabel{$p_1$}{R1}
        \fmflabel{$p_2$}{R2}
      \end{fmfgraph*}
    \end{fmffile}
  \end{gathered} \label{diagram_D1} \\ \nonumber
\end{align}
The expression shall be calculated with the scalar field momenta placed on the mass shell ($p_1^2=p_2^2=m^2$) in the low-energy limit. Since the graviton weakly interacts with a distant observer, it is placed off-shell and related with $p_1$ and $p_2$ by the momentum conservation: $ k^2 = 2\, m^2 + 2 \, p_1\cdot p_2$. This relation allows one to express all scalar products in the amplitude via $m^2$ and $k^2$ alone.

The complete expression for the diagram contains more than $5000$ terms. However, all the loop integrals in the expression are reduced to three Passarino-Veltman integrals:
\begin{align}
    \begin{split}
        A_0 (m_1^2) &\overset{\text{def}}{=} \cfrac{\mu^{4-d}}{i\,\pi^2} \int \cfrac{d^d l}{(2\pi)^d} \,\cfrac{1}{l^2 - m_1^2} \, ,\\
        B_0 (p_1^2, m_1^2, m_2^2) & \overset{\text{def}}{=} \cfrac{\mu^{4-d}}{i\,\pi^2}\int \cfrac{d^d l}{(2\pi)^d} \,\cfrac{1}{l^2 - m_1^2}\,\cfrac{1}{(p_1-l)^2 - m_2^2} \,, \\
        C_0 (p_1^2, p_2^2, (p_1-p_2)^2, m_1^2, m_2^2, m_3^2) & \overset{\text{def}}{=} \cfrac{\mu^{4-d}}{i\,\pi^2} \,\, \int\cfrac{d^d l}{(2\pi)^2} \, \cfrac{1}{(p_1-l)^2 - m_1^2} \, \cfrac{1}{l^2 - m_2^2}\,\cfrac{1}{(p_2-l)^2-m_3^2} \,.
    \end{split}
\end{align}
Here $p_i$ and $m_i$ are arbitrary momenta and masses. Expressions for these integrals are highly complicated, but we evaluate them with the \texttt{Package-X} \cite{Patel:2015tea,Patel:2016fam} and \texttt{FeynHelpers} \cite{Shtabovenko:2016whf}. 

The expression \eqref{diagram_D1} contains only three loop integrals:
\begin{align}
    B_0 (m^2, 0, 0), B_0 (k^2, 0, 0), C_0 (m^2, m^2, k^2, 0, 0, 0).
\end{align}
The first integral is irrelevant since it does not contain the graviton momentum and cannot introduce non-analytic functions of the graviton momentum in the expressions. The second integral does depend on the graviton momentum. The expression for the integral in dimensional regularisation reads
\begin{align}
    B_0 (k^2, 0, 0) = \cfrac{1}{d-4} - \gamma +2 - \ln\left( - \cfrac{\pi\, k^2}{\mu^2} \right),
\end{align}
where $\gamma$ is the Euler's constant and $\mu$ is the renormalisation scale. In the low energy limit, the graviton four-momentum keeps only the spacial part $k^2 = - \left(\vec{k}\right)^2$, which approaches zero. Consequently, the integral has the following leading low-energy behaviour:
\begin{align}
    B_0(k^2, 0, 0) \overset{\text{low-energy}}{\approx} - \ln \left(\vec{k}\right)^2 .
\end{align}
The last integral is equal to the following:
\begin{align}
    \begin{split}
        C_0 (m^2, m^2, k^2, 0, 0, 0) = -\cfrac{2}{ \sqrt{k^2 (k^2 - 4 m^2)} }\Bigg[&  \operatorname{Li}_2 \cfrac{-k^2 + 2 m^2 - \sqrt{k^2 (k^2 - 4 m^2)} }{2\,m^2} \\
        & - \operatorname{Li}_2 \cfrac{-k^2 + 2 m^2 + \sqrt{k^2 (k^2 - 4 m^2)} }{2\,m^2} ~ \Bigg].
     \end{split}
\end{align}
Here $\operatorname{Li}_2$ is the dilogarithm. In the low energy limit, the expression has the following behaviour:
\begin{align}
    C_0 (m^2, m^2, k^2, 0, 0, 0) \overset{\text{low-energy}}{\approx} - \cfrac{1}{m^2}\,\ln \left(\vec{k}\right)^2 .
\end{align}

These relations for the Passarino-Veltman integrals allow one to calculate the low-energy limit of the diagram \eqref{diagram_D1}:
\begin{align}
  \nonumber \\
  \begin{gathered}
    \begin{fmffile}{D2}
      \begin{fmfgraph*}(30,30)
        \fmfleft{L}
        \fmfright{R1,R2}
        \fmf{dbl_wiggly,tension=2}{L,VL}
        \fmf{dashes,tension=2}{R1,V1}
        \fmf{dashes,tension=2}{R2,V2}
        \fmf{dbl_wiggly,tension=0.5}{VL,V1,V2,VL}
        \fmfdot{VL}
        \fmfv{decoration.shape=circle,decoration.filled=empty,decoration.size=5}{V1,V2}
        \fmflabel{$\mu\nu$}{L}
        \fmflabel{$p_1$}{R1}
        \fmflabel{$p_2$}{R2}
      \end{fmfgraph*}
    \end{fmffile}
  \end{gathered} \to i\,\pi^2 \, \kappa \, \left( \cfrac{m}{\mathcal{E}} \right)^6 \cfrac{1}{384} \,\ln \left(\vec{k}\right)^2 \,   k_\mu k_\nu \,. \\ \nonumber
\end{align}
The procedures described in the previous section produce the following expression for the metric
\begin{align}\label{metric_linear_coupling}
    \begin{split}
        g_{00} &= 1 + \left[ - \cfrac{G_\text{N} \,m}{r} -\cfrac{\pi}{6144}\,\cfrac{G_\text{N}\,m^5}{\mathcal{E}^6}\,\cfrac{1}{r^3}  +\, \mathcal{O}\left( \cfrac{1}{\mathcal{E}^7} \right) \right] + \mathcal{O}\left( G_\text{N}^2\right) \,,\\
        g_{0i} &= 0 \, , \\
        g_{ij} &= -\delta_{ij} \left\{ 1 + \left[ \cfrac{G_\text{N}\,m}{r} + \cfrac{\pi}{12288}\,\cfrac{G_\text{N}\,m^5}{\mathcal{E}^6}\,\cfrac{1}{r^3} +\, \mathcal{O}\left( \cfrac{1}{\mathcal{E}^7} \right) \right] + \mathcal{O}\left( G_\text{N}^2\right) \right\} ,
    \end{split}
\end{align}
and for the two-body potential
\begin{align}\label{V_linear_coupling}
    V = - \cfrac{G_\text{N}\,m_1\,m_2}{r} \,\left[ 1 - \cfrac{\pi^2}{96}\,\cfrac{m_1^4+m_2^4}{\mathcal{E}^6} \, \cfrac{1}{r^2}  +\, \mathcal{O}\left( \cfrac{1}{\mathcal{E}^7} \right)  \right] + \mathcal{O}\left( G_\text{N}^2\right)\,.
\end{align}
The expression for the metric is given in the Cartesian frame, and $m$ is the mass of a particle that is non-minimally coupled to the Gauss-Bonnet term. In the expression for the two-body potential, $m_1$ and $m_2$ are masses of two scalars.

We proceed with the quadratic coupling case. In terms of parametrisation \eqref{the_action} the coupling function reads
\begin{align}\label{the_quadratic_coupling_definition}
  f(\varphi) &= \alpha \,\varphi^2 .
\end{align}
Here $\alpha$ is the coupling with the mass dimension $-2$. In terms of parametrisation \eqref{the_action_2}, the coupling to the Gauss-Bonnet term reads
\begin{align}
    \int d^4 x \, \sqrt{-g} \left[ - \cfrac{1}{\kappa^2 \, \mathcal{E}^2} ~ \cfrac{\varphi^2}{\mathcal{E}^2} ~ \mathcal{G} \right] \,
\end{align}
where the energy scale $\mathcal{E}$ is related with the coupling:
\begin{align}
  \mathcal{E} = \cfrac{1}{ \sqrt{ \kappa \sqrt{\alpha}} } \,.
\end{align}

Following the same hierarchy reasoning, we account for the leading order corrections in the standard and the non-minimal couplings. Consequently, the matrix element describing the low-energy limit reads:
\begin{align}
  \nonumber \\
  \begin{gathered}
    \begin{fmffile}{Series_3_D1}
      \begin{fmfgraph*}(30,30)
        \fmfleft{L}
        \fmfright{R1,R2}
        \fmf{dbl_wiggly}{L,V}
        \fmf{dashes}{R1,V,R2}
        \fmfv{decoration.shape=circle,decoration.size=15pt,decoration.filled=shaded}{V}
        \fmflabel{$\mu\nu,k$}{L}
        \fmflabel{$p_1$}{R1}
        \fmflabel{$p_2$}{R2}
      \end{fmfgraph*}
    \end{fmffile}
  \end{gathered}
  \hspace{10pt} =& ~ \Bigg[
  \begin{gathered}
    \begin{fmffile}{Series_3_D2}
      \begin{fmfgraph}(30,30)
        \fmfleft{L}
        \fmfright{R1,R2}
        \fmf{dbl_wiggly}{L,V}
        \fmf{dashes}{R1,V,R2}
        \fmfdot{V}
      \end{fmfgraph}
    \end{fmffile}
  \end{gathered}
  + ~
  \begin{gathered}
    \begin{fmffile}{Series_3_D3}
      \begin{fmfgraph}(30,30)
        \fmfleft{L}
        \fmfright{R1,R2}
        \fmf{dashes,tension=2}{R1,VR,R2}
        \fmf{phantom,tension=0.7}{L,VR}
        \fmffreeze
        \fmf{dbl_wiggly}{L,V}
        \fmf{phantom}{V,VR}
        \fmffreeze
        \fmf{dbl_wiggly,right=1}{V,VR,V}
        \fmfdot{V}
        \fmfv{decoration.shape=circle,decoration.filled=empty,decoration.size=5}{VR}
      \end{fmfgraph}
    \end{fmffile}
  \end{gathered}
  + ~
  \begin{gathered}
    \begin{fmffile}{Series_3_D4}
      \begin{fmfgraph}(30,30)
        \fmfleft{L}
        \fmfright{R1,R2}
        \fmf{dbl_wiggly,tension=2}{L,V}
        \fmf{dashes,tension=0.5}{R1,V}
        \fmf{dashes,tension=0.5}{R2,V}
        \fmffreeze
        \fmf{phantom,tension=2}{R1,V1}
        \fmf{phantom}{V1,V}
        \fmffreeze
        \fmf{dbl_wiggly,left=0.7}{V,V1}
        \fmfdot{V1}
        \fmfv{decoration.shape=circle,decoration.filled=empty,decoration.size=5}{V}
      \end{fmfgraph}
    \end{fmffile}
  \end{gathered}
  + ~
  \begin{gathered}
    \begin{fmffile}{Series_3_D5}
      \begin{fmfgraph}(30,30)
        \fmfleft{L}
        \fmfright{R1,R2}
        \fmf{dbl_wiggly,tension=2}{L,V}
        \fmf{dashes,tension=0.5}{R1,V}
        \fmf{dashes,tension=0.5}{R2,V}
        \fmffreeze
        \fmf{phantom,tension=2}{R2,V1}
        \fmf{phantom}{V1,V}
        \fmffreeze
        \fmf{dbl_wiggly,right=0.7}{V,V1}
        \fmfdot{V1}
        \fmfv{decoration.shape=circle,decoration.filled=empty,decoration.size=5}{V}
      \end{fmfgraph}
    \end{fmffile}
  \end{gathered}
  +\mathcal{O}\left(\mathcal{E}^{-5}\right)\Bigg]+ \okappa{3} .
  \nonumber
\end{align}

In full analogy with the previous case, the expression contains many terms. We analyse each contribution to the matrix element and study their low-energy limit. First and foremost, two of these diagrams are irrelevant for the low-energy. The following diagrams do have loop integrals, but they involve only the momenta of scalars, which are fixed on the shell and do not contribute to the low-energy limit
\begin{align}
  \nonumber \\
  \begin{gathered}
    \begin{fmffile}{Series_4_D1}
      \begin{fmfgraph}(30,30)
        \fmfleft{L}
        \fmfright{R1,R2}
        \fmf{dbl_wiggly,tension=2}{L,V}
        \fmf{dashes,tension=0.5}{R1,V}
        \fmf{dashes,tension=0.5}{R2,V}
        \fmffreeze
        \fmf{phantom,tension=2}{R1,V1}
        \fmf{phantom}{V1,V}
        \fmffreeze
        \fmf{dbl_wiggly,left=0.7}{V,V1}
        \fmfdot{V1}
        \fmfv{decoration.shape=circle,decoration.filled=empty,decoration.size=5}{V}
      \end{fmfgraph}
    \end{fmffile}
  \end{gathered}
  ~ , ~
  \begin{gathered}
    \begin{fmffile}{Series_4_D2}
      \begin{fmfgraph}(30,30)
        \fmfleft{L}
        \fmfright{R1,R2}
        \fmf{dbl_wiggly,tension=2}{L,V}
        \fmf{dashes,tension=0.5}{R1,V}
        \fmf{dashes,tension=0.5}{R2,V}
        \fmffreeze
        \fmf{phantom,tension=2}{R2,V1}
        \fmf{phantom}{V1,V}
        \fmffreeze
        \fmf{dbl_wiggly,right=0.7}{V,V1}
        \fmfdot{V1}
        \fmfv{decoration.shape=circle,decoration.filled=empty,decoration.size=5}{V}
      \end{fmfgraph}
    \end{fmffile}
  \end{gathered} \to A_0(m^2) , B_0(p_1^2,0,m^2), B_0 (p_2^2,0,m^2). \\ \nonumber
\end{align}

The only remaining diagram describes the leading order corrections and its expression is compact enough to be presented in print:
\begin{align}
  \nonumber \\
  \begin{gathered}
    \begin{fmffile}{Scalar_9_Detail}
      \begin{fmfgraph*}(40,40)
        \fmfleft{L}
        \fmfright{R1,R2}
        \fmf{dashes,tension=2}{R1,VR,R2}
        \fmf{phantom,tension=0.7}{L,VR}
        \fmffreeze
        \fmf{dbl_wiggly}{L,V}
        \fmf{phantom}{V,VR}
        \fmffreeze
        \fmf{dbl_wiggly,right=1}{V,VR,V}
        \fmfdot{V}
        \fmfv{decoration.shape=circle,decoration.filled=empty,decoration.size=5}{VR}
        \fmflabel{$\mu\nu$, $k$}{L}
        \fmflabel{$p_1$}{R1}
        \fmflabel{$p_2$}{R2}
      \end{fmfgraph*}
    \end{fmffile}
  \end{gathered}
  = \kappa\, \cfrac{1}{\mathcal{E}^4} \,\cfrac{1}{64}\,k^6\,i\,\pi^2\, B_0(k^2,0,0) \, \Bigg[& 2 (d^4-15 d^3+88 d^2-220 d+198) \eta_{\mu\nu} \\
    & - \cfrac{d^5-13 d^4+66 d^3-100 d^2-90 d+252}{d-1}\,\theta_{\mu\nu}(k) \Bigg]. \nonumber
\end{align}
Here $d$ is the number of space-time dimension, and $\theta$ is the standard gauge projector:
\begin{align}
  \theta_{\mu\nu}(k) = \eta_{\mu\nu} - \cfrac{k_\mu\,k_\nu}{k^2}\,.
\end{align}
As discussed above, the only present Passarino-Veltman integral has the following low-energy asymptotic:
\begin{align}
  B_0 \left( k^2, 0, 0 \right) \overset{\text{low-energy}}{\approx} - \ln \abs{\vec{k}}^2.
\end{align}

Consequently, the procedure described in the previous section produces the following expression for the metric
\begin{align}\label{metric_quadratic_coupling}
    \begin{split}
        g_{00} &= 1 + \left[ -\cfrac{G_\text{N} \,m}{r} +  \cfrac{165\,\pi}{32}\,\cfrac{G_\text{N}}{\mathcal{E}^4\,m}\,\cfrac{1}{r^7} + \cfrac{525\,\pi}{4}\,\cfrac{G_\text{N}}{\mathcal{E}^4\,m^3}\,\cfrac{1}{r^9} + \mathcal{O}\left(\cfrac{1}{\mathcal{E}^5}\right)  \right] + \mathcal{O} \left( G_\text{N}^2 \right)\,, \\
        g_{0i} &= 0 \, , \\
        g_{ij} &= -\delta_{ij}\left\{ 1 + \left[ \cfrac{G_\text{N}}{r} - \cfrac{165\,\pi}{64}\,\cfrac{G_\text{N}}{\mathcal{E}^4} \,\cfrac{1}{r^7} + \cfrac{2625\,\pi}{16}\, \cfrac{G_\text{N}}{\mathcal{E}^4\,m^3}\,\cfrac{1}{r^9} + \mathcal{O}\left(\cfrac{1}{\mathcal{E}^5}\right) \right] + \mathcal{O} \left( G_\text{N}^2 \right) \right\}\, .
    \end{split}
\end{align}
and for the two-body interaction potential
\begin{align}\label{V_quadratic_coupling}
  V = - \cfrac{G_\text{N}\,m_1\,m_2}{r} \left[ 1 + 65\,\pi^2\, \cfrac{m_1^2+m_2^2}{\mathcal{E}^4\,m_1^2\,m_2^2}\,\cfrac{1}{r^6} + 2730\,\pi^2\, \cfrac{1}{\mathcal{E}^4\,m_1^2\,m_2^2}\, \cfrac{1}{r^8}  + \mathcal{O}\left(\cfrac{1}{\mathcal{E}^6}\right) \right] + \mathcal{O}\left(G_\text{N}^2 \right) .
\end{align}
The expression for the metric is given in the Cartesian frame, and $m$ is the mass of a scalar particle that is non-minimally coupled to the Gauss-Bonnet term. In the expression for the two-body potential, $m_1$ and $m_2$ are masses of scalars. Finally, these expressions contain higher powers of $1/r$. They appear because of the following terms in the expressions:
\begin{align}
  \int \cfrac{d^3 \vec{k}}{(2\pi)^3}\, \abs{\vec{k}}^4 \ln \abs{\vec{k}}^2 & = \cfrac{5! }{2\,\pi\,r^7}\,, & \int \cfrac{d^3 \vec{k}}{(2\pi)^3}\, \abs{\vec{k}}^6 \ln \abs{\vec{k}}^2 & = \cfrac{7!}{2\,\pi\,r^9}\, .
\end{align}
The corresponding Fourier transform is discussed in the Appendix \ref{Appendix_Fourier}.

We return to discussing these results in Section \ref{Section_conclusion}. Here, we point to the two most important features. First, the same models were considered within the standard PPN formalism \cite{Sotiriou:2006pq,Rannu:2012fgy,will2018theory,Will:2014kxa}. It was found that the Gauss-Bonnet term contribution is strongly suppressed and does not contribute to the PPN metric. Our results show that the Gauss-Bonnet contribution appears in higher powers of $1/r$, so they do not enter the PPN metric. Secondly, this strong suppression of the Gauss-Bonnet coupling makes detecting it challenging. The coupling may influence light scattering on a massive object, such as the small-angle scattering. The growing precision of measurement may provide enough data to constrain the magnitude of the discussed corrections. We discuss this opportunity in the following sections.

\section{Classical Scattering}\label{Section_classical_scattering}

The previous section shows that the non-minimal coupling to the Gauss-Bonnet term changes the gravitational potential of a point particle. As discussed in Section \ref{Section_perturbative_quantum_gravity}, within general relativity, the gravitational field of a point-like particle matches the black hole metric in a few leading orders. Consequently, one can use metrics \eqref{metric_linear_coupling} and \eqref{metric_quadratic_coupling} to describe the leading contribution of the non-minimal Gauss-Bonnet coupling to the black hole metric. In turn, we can probe these metrics with the motion of test particles, for instance, by the small-angle light scattering.

This section reviews the classical results on the small-angle light scattering. The motion of a particle in the external gravitational field of a black hole, including particle scattering, is well understood and discussed in many textbooks and publications \cite{Chandrasekhar:1985kt,Misner:1973prb, Perlick:2021aok,Collins:1973xf}. The next section will compare these classical results with the quantum case. Analyzing the scattering cross-section in the classical case allows us to study the contribution of the non-minimal Gauss-Bonnet coupling to the small-angle scattering.

The scattering problem is stated as follows. At the initial moment, the particle approaches the black hole from infinity. The particle is displaced from the black hole centre by the distance $b$, which is called the impact parameter. Using the spherical frame where the particle's initial angular coordinate is zero is more convenient. Consequently, using the particle angular velocity $dr/d\varphi$ is more appropriate. In the initial moment, the angular velocity is negative because the angular coordinate grows, but the radial coordinate is reduced as the particle approaches the black hole.

We study only the scattering of a particle and do not address situations where a particle is captured by or orbits the black hole. When a particle is scattered by a black hole, it initially moves towards the black hole, reaches a minimum distance, and then starts moving away from the black hole. This minimum distance is known as the turnaround radius $r_\text{turnaround}$ because, at this point, the particle changes the direction of its angular velocity. When a particle moves away from the black hole, it enters the asymptotically flat region where one can calculate the scattering angle $\chi$ on which the particle was scattered.

The scattering angle depends on the impact parameter, but it is more helpful to operate with the dependence of the impact parameter on the scattering angle: $b = b(\chi)$. To solve the scattering problem, we need to find the dependence of the scattering angle on the impact parameter. Analytical solutions to the scattering problem are known for a few exceptional cases. We use the small-angle scattering to study the problem without constructing numerical or analytic solutions. We assume that the scattering angle $\chi$ is small and expand the impact parameter and all other quantities in a Taylor series with respect to small $\chi$.

To proceed with the derivation, we obtain the equation describing the motion of light in a spherical symmetric gravitational field. The following Lagrangian describes the motion of a relativistic particle in curved spacetime:
\begin{align}
  \mathcal{L} = -\cfrac12\, g_{\mu\nu} \, \cfrac{dx^\mu}{d\tau} \cfrac{dx^\mu}{d\tau}\,.
\end{align}
Here, $x^\mu=x^\mu(\tau)$ denotes the particle coordinates, and $\tau$ is the affine parameter. For the metric ansatz with the spherical symmetry, the Lagrangian reads:
\begin{align}\label{particle_Lagrangian}
  \mathcal{L} = \cfrac12\, A\,\left( t' \right)^2 - \cfrac12\,\cfrac{1}{B} \,\left( r' \right)^2 -\cfrac12\,r^2 \left( \theta' \right)^2 - \cfrac12\,r^2 \sin^2\theta \left( \varphi' \right)^2.
\end{align}
Here $x^\mu = \left( t(\tau), r(\tau), \theta(\tau), \varphi(\tau) \right)$ are coordinates of the particle, the prime $'$ denotes a derivative with respect to the affine parameter $\tau$, and $A$, $B$ are metric functions that depend only on $r$. This Lagrangian admits the standard Euler-Lagrange equations:
\begin{align}
  \cfrac{\delta\mathcal{L}}{\delta q} - \cfrac{d}{d\tau} \,\cfrac{\delta\mathcal{L}}{\delta q'} =0.
\end{align}
First, since the Lagrangian does not explicitly depend on $t$ and $\varphi$, the corresponding Euler-Lagrange equations are conservation laws of the following invariants:
\begin{align}
  E & \overset{\text{def}}{=} A \, t' \,, & L & \overset{\text{def}}{=} r^2 \,\sin^2\theta\,\varphi' .
\end{align}
The first one describes the particle energy, and the second one describes the angular momentum. Second, the Euler-Lagrange equations admit $\theta=\pi/2$ as a solution corresponding to motion in a single plane. Consequently, the Lagrangian \eqref{particle_Lagrangian} reduces to the following effective one-dimensional form:
\begin{align}
  \mathcal{L} = \cfrac12 \left[ \cfrac{E^2}{A} - \cfrac{\left( r' \right)^2}{B} - \cfrac{L^2}{r^2} \right] .
\end{align}
There is a better way to proceed instead of operating with the corresponding Euler-Lagrange equation. By construction, the Lagrangian is related to a square of the particle four-velocity $u^\mu$:
\begin{align}
  \mathcal{L} &= \cfrac12 \,g_{\mu\nu} \, u^\mu \,u^\nu, & u^\mu & = \cfrac{d x^\mu}{d\tau} \, .
\end{align}
Therefore, the Lagrangian vanishes for the massless case:
\begin{align}
  \cfrac{E^2}{A} - \cfrac{\left( r' \right)^2}{B} - \cfrac{L^2}{r^2} =0 .
\end{align}
The equation reduces to a single first-order differential equation:
\begin{align}
  \left(\cfrac{dr}{d\tau}\right)^2 = B \left[ \cfrac{E^2}{A} - \cfrac{L^2}{r^2} \right] \,.
\end{align}
It is more useful to operate with the angular velocity $dr/d\varphi$. The conservation of angular momentum provides a way to change the motion parameter:
\begin{align}
  \cfrac{d r}{ d\varphi} = \cfrac{ \frac{d r}{d \tau} }{ \frac{d\varphi}{d\tau}} = \cfrac{ r' }{L/r^2} \, .
\end{align}
This gives the following equation
\begin{align}\label{orbital_equation}
  \left( \cfrac{d r}{ d\varphi} \right)^2 = r^2\, B \, \left[ \cfrac{1}{A} \, \cfrac{r^2}{\left(L/E\right)^2}- 1 \right].
\end{align}
For the sake of brevity, we will refer to this equation as the orbital equation.

Before the particle reaches the turnaround radius, one can integrate the orbital equation \eqref{orbital_equation} by choosing the negative sign of the square root:
\begin{align}
  \int\limits_{r(\varphi)}^\infty\, \cfrac{dr}{ r} ~ \cfrac{1}{\sqrt{ B(r) \, \left[ \cfrac{1}{A(r)} \, \cfrac{r^2}{\left(L/E\right)^2}- 1 \right] } } = \varphi .
\end{align}
The equation simplifies since $L/E$ is the impact parameter $b$:
\begin{align}\label{orbital_equation_integrated}
  \int\limits_{r(\varphi)}^\infty \cfrac{dr}{ r} \, \cfrac{1}{\sqrt{ B(r) \, \left[ \cfrac{1}{A(r)} \, \left( \cfrac{r}{b} \right)^2- 1 \right] } } = \varphi .
\end{align}

The next step is to find the turnaround radius. By definition, when a particle reaches the turnaround radius, its angular velocity $dr/d\varphi$ becomes zero, and the right-hand side of the orbital equation \eqref{orbital_equation} vanishes. The previous results \cite{Papageorgiou:2022umj} show that $B(r) > 0$ outside the horizon, so the outermost root of the following expression gives $r_\text{turbaround}$:
\begin{align}\label{turnaround_radius_equation}
  \cfrac{1}{A(r)} \left( \cfrac{r}{b} \right)^2 - 1 = 0.
\end{align}

Since the spacetime admits the flat asymptotic, the scattering angle $\chi$ is related to the angular coordinate $\varphi_\text{turnaround}$ at which the particle reaches the turnaround radius:
\begin{align}
  \chi + 2 \, \varphi_\text{turnaround} = \pi.
\end{align}
Consequently, one obtains the following formula relating the scattering angle and the impact parameter:
\begin{align}
  \chi(b) = \abs{ \pi - 2 \int\limits_{r_\text{turnaround}}^\infty \frac{dr}{ r\, \sqrt{ B(r) \, \left[ \frac{1}{A} \, \left( \frac{r}{b} \right)^2 - 1\right] } } } .
\end{align}

Finally, this provides a way to calculate the differential cross section $d\sigma$, which shows what part of an incoming particle stream will reach space infinity at the solid angle $d\Omega=\sin\chi \, d\chi \, d\varphi$. The formula relating the differential cross section to the impact parameter $b(\chi)$ is known:
\begin{align}\label{the_classical_differential_cross-section}
  d\sigma = \cfrac{b(\chi)}{\sin\chi} \abs{ \cfrac{d\, b(\chi)}{d\chi} } d\Omega \, .
\end{align}

\begin{figure}[htbp]
  \centering
  \begin{minipage}{0.49\linewidth}
    \centering
    \includegraphics[width=\linewidth]{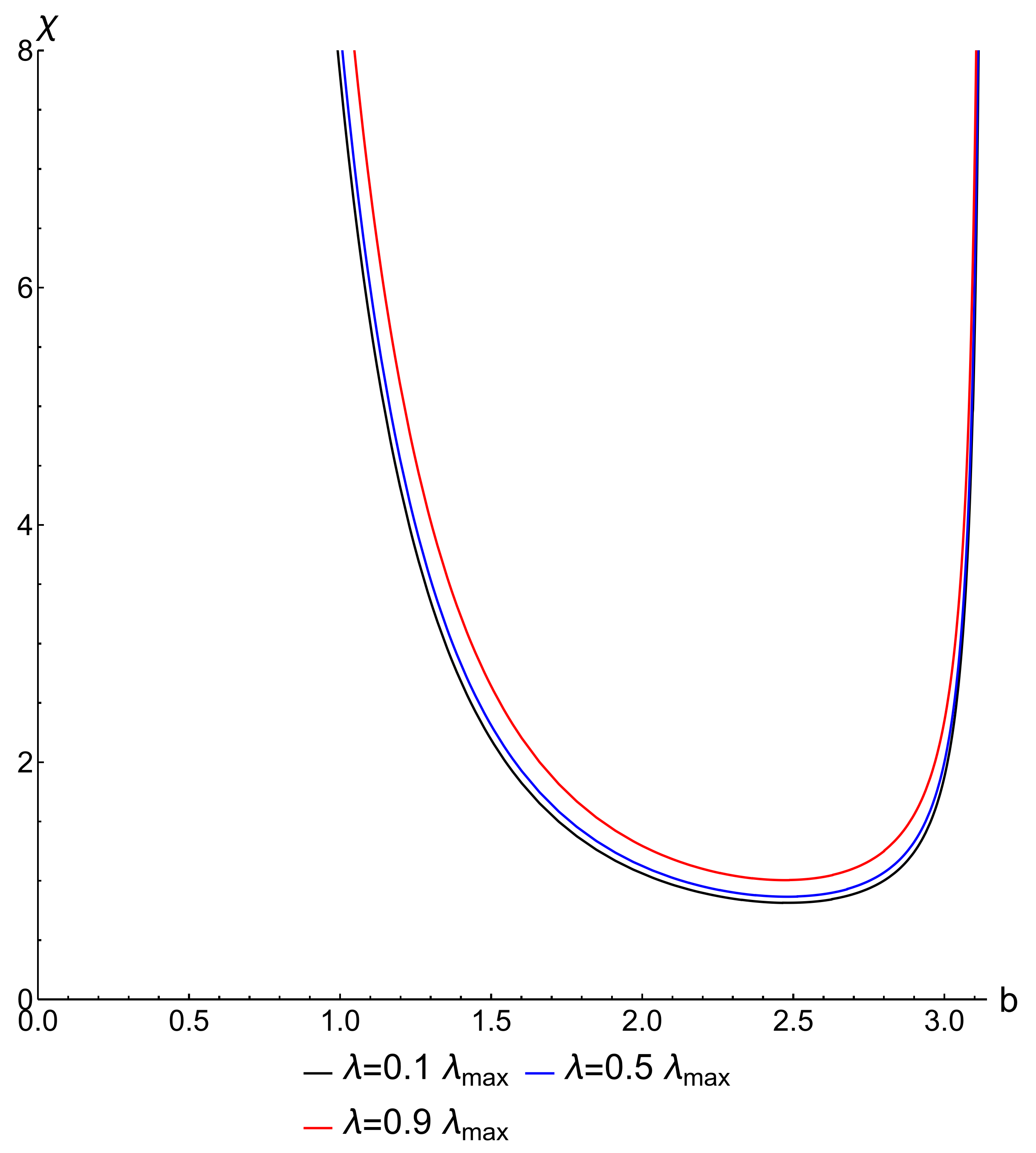}
  \end{minipage}
  \caption{Differential scattering cross-section for various numerical solutions for linear coupling}\label{fig:scattering_linear_coupling}
\end{figure}

\begin{figure}[htbp]
  \centering
  \begin{minipage}{0.49\linewidth}
    \centering
    \includegraphics[width=\linewidth]{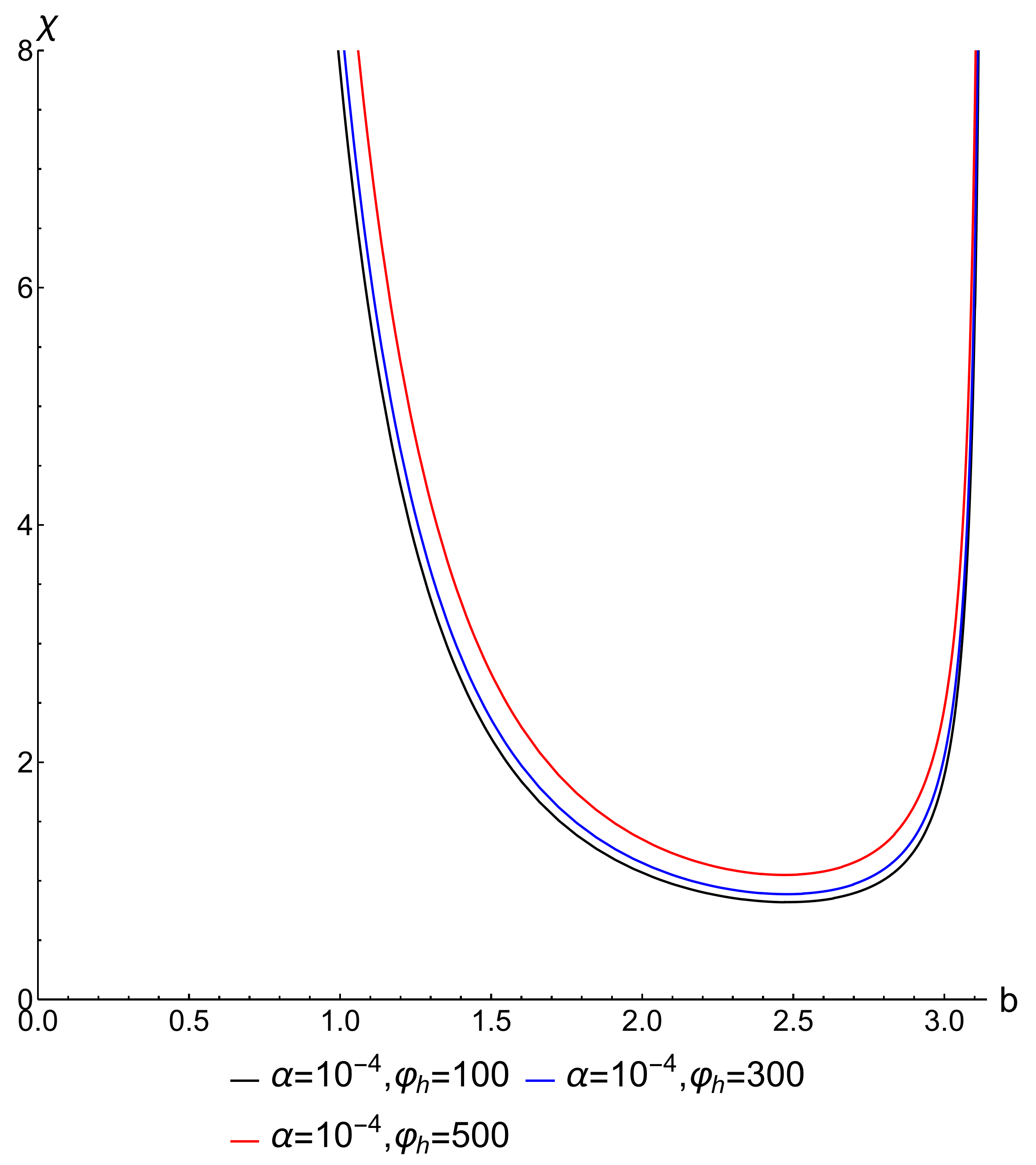}
  \end{minipage}
  \begin{minipage}{0.49\linewidth}
    \centering
    \includegraphics[width=\linewidth]{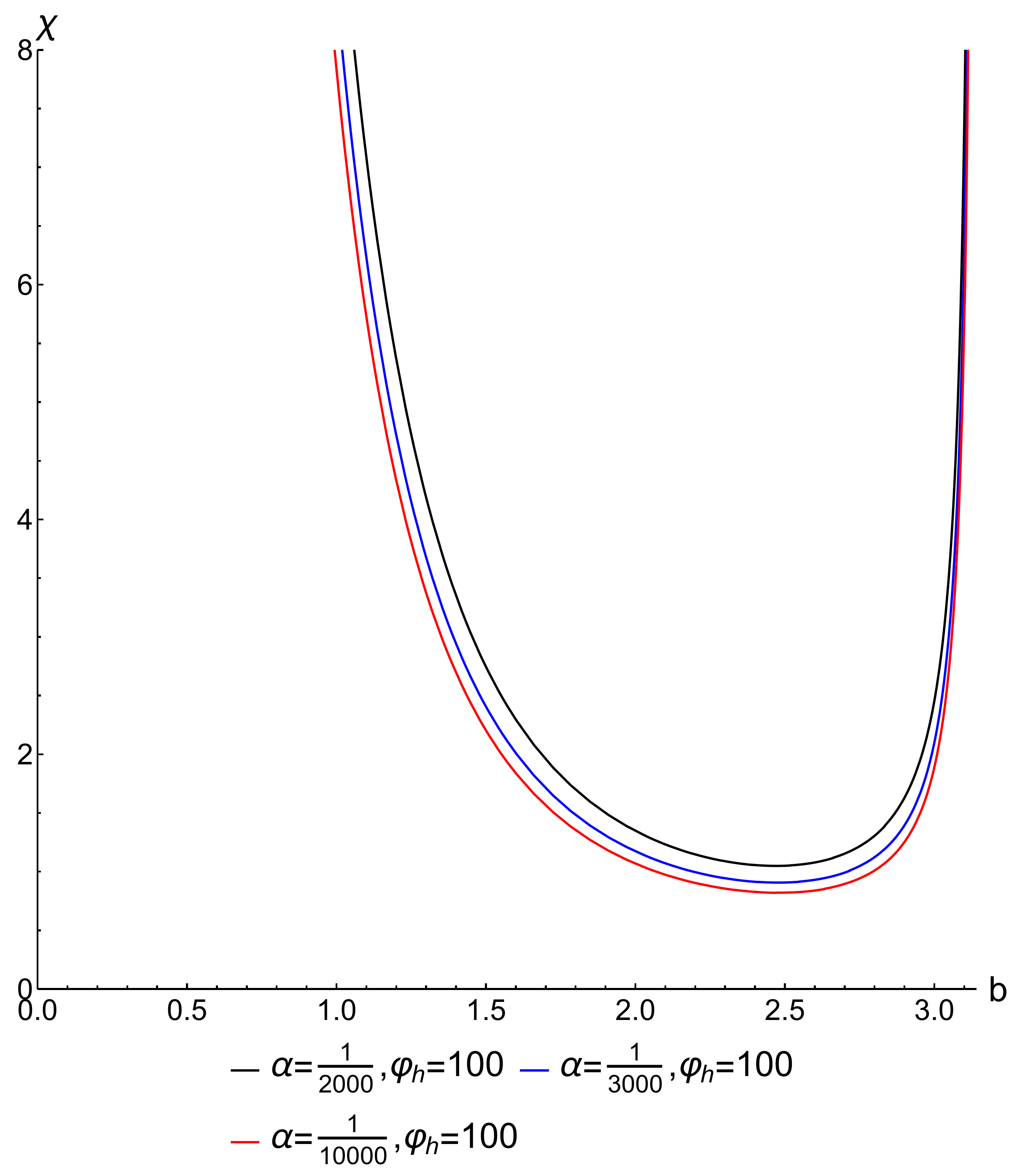}
  \end{minipage}
  \caption{Differential scattering cross-section for various numerical solutions for quadratic coupling}\label{fig:scattering_quadratic_coupling}
\end{figure}

The formula provides a tool for constructing classical (relativistic, not quantum) scattering cross-sections. We present some cross-sections calculated for the numerical solutions discussed in Section \ref{Section_black_hole_metric} in Figures \ref{fig:scattering_linear_coupling}, \ref{fig:scattering_quadratic_coupling}. The model admits scalar field shift symmetry for the linear coupling case, so the cross-section is independent of the scalar field value at the horizon. For that case, we present a single plot which shows how the cross-section depends on the coupling. The model does not admit the shift symmetry for the quadratic coupling case, so the scattering is defined by both the coupling and the scalar field value at the horizon. Consequently, we present two plots showing these dependencies.

The differential cross-section \eqref{the_classical_differential_cross-section} provides a tool to compare classical and quantum scattering, but it faces two challenges. First and foremost, the analytic expression for $b(\chi)$ is exceptionally challenging to obtain. To our knowledge, even for the Schwarzschild black hole, the analytic expression for $b(\chi)$ was never obtained. Secondly, it is challenging to introduce the notion of the scattering angle within quantum field theory since particles are subjected to the uncertainty principle. 

We overcome the first challenge by expanding $b(\chi)$ in a power series for small $\chi$:
\begin{align}
  b(\chi) = 4 \, G_\text{N}\,M\,\left[ \cfrac{b_{-1}}{\chi} + b_0 + b_1 \chi + \mathcal{O}\left(\chi^2\right) \right].
\end{align}
In this expression, $b_{-1}$, $b_0$, $b_1$, $\cdots$ are numerical coefficients. The presence of $\chi^{-1}$ term is necessary for the scattering angle singular behaviour. Since gravity is a long-range force, no matter how far a particle is from a black hole, it experiences gravitational attraction. Consequently, for any large but finite impact parameter $b$, the scattering angle $\chi$ will be small but non-zero and vanish only in the $b\to\infty$ limit. The scattering angle expansion gives the following expression for the differential cross-section:
\begin{align}
  \cfrac{d\sigma}{d\Omega} = \cfrac{b(\chi)}{\sin\chi} \abs{\cfrac{d b(\chi)}{d \chi}} = (4\,G_\text{N}\,M)^2\, \left[ \cfrac{b_{-1}^2}{\chi^4} + \cfrac{b_{-1}\,b_0}{\chi^3} + \cfrac{1}{6} \,\cfrac{b_{-1}^2}{\chi^2} +\left[ \cfrac16\,b_{-1}\,b_0 - b_0\,b_1 - b_{-1}\,b_2 \right] \cfrac{1}{\chi} + \mathcal{O}\left( \chi^0 \right) \right].
\end{align}

We completely negate the issue with the scattering angle by focusing on the differential cross-section rather than the scattering angle. To be exact, one can introduce the notion of the scattering angle using the eikonal approximation. In that case, the perturbative quantum gravity recovers the general relativity result for the small-angle scatterings \cite{Bjerrum-Bohr:2014zsa,Bai:2016ivl,Chi:2019owc}. Our perturbative quantum gravity calculations for general relativity presented in the next section give the following expression for the cross-section at the leading order
\begin{align}
    \cfrac{d\sigma}{d\Omega} = (4\,G_\text{N}\,M)^2\Bigg[ \cfrac{1}{\chi^4} + \cfrac{1}{3!}\,\cfrac{1}{\chi^2} + \cfrac{11}{6!} + \cfrac{31}{6\times 7!}\, \chi^2 + \cfrac{41}{2\times 9!} \,\chi^4 + \mathcal{O}\left(\chi^5\right)\Bigg] + \mathcal{O}\left(G_\text{N}^3\right) .
\end{align}
Consequently, in the next section, we calculate the cross-section, obtain corrections generated by the non-minimal Gauss-Bonnet coupling, and study whether they are relevant to small-angle scattering.

\section{Quantum scattering}\label{Section_quantum_scattering}

In general relativity, the scattering of two scalars with different masses $m_1$ and $m_2$ that are minimally coupled to gravity is described by the following matrix element:
\begin{align}
  \begin{split}
    \\
    \begin{gathered}
      \begin{fmffile}{Series_5_D1}
        \begin{fmfgraph*}(30,30)
          \fmfleft{L1,L2}
          \fmfright{R1,R2}
          \fmf{dashes}{L1,V}
          \fmf{dashes}{L2,V}
          \fmf{dashes}{R1,V}
          \fmf{dashes}{R2,V}
          \fmfv{decoration.shape=circle,decoration.size=15pt,decoration.filled=shaded}{V}
          \fmflabel{$p_1$,$m_1$}{L1}
          \fmflabel{$p_3$,$m_1$}{L2}
          \fmflabel{$p_2$,$m_2$}{R1}
          \fmflabel{$p_4$,$m_2$}{R2}
        \end{fmfgraph*}
      \end{fmffile}
    \end{gathered}
    \hspace{30pt} =& ~
    \begin{gathered}
      \begin{fmffile}{Series_5_D2}
        \begin{fmfgraph}(30,30)
          \fmfbottom{B1,B2}
          \fmftop{T1,T2}
          \fmf{dashes}{B1,V1,T1}
          \fmf{dashes}{B2,V2,T2}
          \fmf{dbl_wiggly}{V1,V2}
          \fmfdot{V1,V2}
        \end{fmfgraph}
      \end{fmffile}
    \end{gathered}\\
    & + ~
    \begin{gathered}
      \begin{fmffile}{Series_5_D3}
        \begin{fmfgraph}(30,30)
          \fmfbottom{B1,B2}
          \fmftop{T1,T2}
          \fmf{dashes}{B1,V1,T1}
          \fmf{dashes}{B2,V2,T2}
          \fmf{dbl_wiggly}{V1,V2}
          \fmfdot{V1}
          \fmfblob{10}{V2}
        \end{fmfgraph}
      \end{fmffile}
    \end{gathered}
    ~ + ~
    \begin{gathered}
      \begin{fmffile}{Series_5_D4}
        \begin{fmfgraph}(30,30)
          \fmfbottom{B1,B2}
          \fmftop{T1,T2}
          \fmf{dashes}{B1,V1,T1}
          \fmf{dashes}{B2,V2,T2}
          \fmf{dbl_wiggly}{V1,V2}
          \fmfdot{V2}
          \fmfblob{10}{V1}
        \end{fmfgraph}
      \end{fmffile}
    \end{gathered}
    ~ + ~
    \begin{gathered}
      \begin{fmffile}{Series_5_D5}
        \begin{fmfgraph}(30,30)
          \fmfbottom{B1,B2}
          \fmftop{T1,T2}
          \fmf{phantom,tension=0.5}{L1,R1,R2,L2,L1}
          \fmf{dashes}{B1,L1}
          \fmf{dashes}{B2,R1}
          \fmf{dashes}{T1,L2}
          \fmf{dashes}{T2,R2}
          \fmfdot{L1,L2,R1,R2}
          \fmffreeze
          \fmf{dashes}{L1,L2}
          \fmf{dashes}{R1,R2}
          \fmf{dbl_wiggly}{L1,R1}
          \fmf{dbl_wiggly}{L2,R2}
        \end{fmfgraph}
      \end{fmffile}
    \end{gathered}
    ~ + ~
    \begin{gathered}
      \begin{fmffile}{Series_5_D6}
        \begin{fmfgraph}(30,30)
          \fmfbottom{B1,B2}
          \fmftop{T1,T2}
          \fmf{phantom,tension=0.5}{L1,R1,R2,L2,L1}
          \fmf{dashes}{B1,L1}
          \fmf{dashes}{B2,R1}
          \fmf{dashes}{T1,L2}
          \fmf{dashes}{T2,R2}
          \fmfdot{L1,L2,R1,R2}
          \fmffreeze
          \fmf{dashes}{L1,L2}
          \fmf{dashes}{R1,R2}
          \fmf{dbl_wiggly}{L1,R2}
          \fmf{dbl_wiggly}{L2,R1}
        \end{fmfgraph}
      \end{fmffile}
    \end{gathered}
    ~ + ~
    \begin{gathered}
      \begin{fmffile}{Series_5_D7}
        \begin{fmfgraph}(30,30)
          \fmfbottom{B1,B2}
          \fmftop{T1,T2}
          \fmf{phantom,tension=.5}{V1,V0,V2}
          \fmf{phantom,tension=.2}{V1,V2}
          \fmf{dashes}{B1,V1}
          \fmf{dashes}{T1,V2}
          \fmf{dashes}{B2,V0,T2}
          \fmfdot{V0,V1,V2}
          \fmffreeze
          \fmf{dashes}{V1,V2}
          \fmf{dbl_wiggly}{V1,V0}
          \fmf{dbl_wiggly}{V2,V0}
        \end{fmfgraph}
      \end{fmffile}
    \end{gathered}
    ~ + ~
    \begin{gathered}
      \begin{fmffile}{Series_5_D8}
        \begin{fmfgraph}(30,30)
          \fmfbottom{B1,B2}
          \fmftop{T1,T2}
          \fmf{phantom,tension=.5}{V1,V0,V2}
          \fmf{phantom,tension=.2}{V1,V2}
          \fmf{dashes}{B2,V1}
          \fmf{dashes}{T2,V2}
          \fmf{dashes}{B1,V0,T1}
          \fmfdot{V0,V1,V2}
          \fmffreeze
          \fmf{dashes}{V1,V2}
          \fmf{dbl_wiggly}{V1,V0}
          \fmf{dbl_wiggly}{V2,V0}
        \end{fmfgraph}
      \end{fmffile}
    \end{gathered}
    ~ + ~
    \begin{gathered}
      \begin{fmffile}{Series_5_D9}
        \begin{fmfgraph}(30,30)
          \fmftop{T1,T2}
          \fmfbottom{B1,B2}
          \fmf{dashes}{B1,V1,T1}
          \fmf{dashes}{B2,V2,T2}
          \fmf{dbl_wiggly,right=1}{V1,V2,V1}
          \fmfdot{V1,V2}
        \end{fmfgraph}
      \end{fmffile}
    \end{gathered}\\
    &
    +\okappa{6}.
  \end{split}
\end{align}
In this particular matrix element, we presented contributions up to the next-to-leading order in the gravitational coupling $\okappa{4}$, and the blob notes the corresponding one-loop vertex function. Since the two graviton exchange processes contribute to the matrix element, the gravitational scattering cannot be reduced to the effects described by the vertex function.

Our goal is to describe the scattering of a massless object on the scalar field non-minimally coupled to the Gauss-Bonnet term. To do this, we introduce an additional scalar field to the model. The new scalar field is massless and only admits minimal coupling to gravity.
\begin{align}
  \begin{split}
    p_1 & = ( \sqrt{M^2 + p^2}, 0, 0, p ) , \\
    p_2 & = ( p, 0, 0, -p ) , \\
    p_3 & = ( \sqrt{M^2 + p^2}, p\sin\chi, 0, p \cos\chi ) , \\
    p_4 & = ( p, -p\sin\chi, 0, -p\cos\chi ) .
  \end{split}
  &
  \begin{cases}
    s = (p_1 + p_2 )^2 = \left(p + \sqrt{M^2 + p^2} \right)^2 \, ,\\
    t = (p_1 + p_3 )^2 = -4\,p^2\, \sin^2\frac{\chi}{2} \, ,\\
    u = (p_1 + p_4 )^2 = 2\,M^2 - s -t \, .
  \end{cases}
\end{align}
Momenta $p_1$ and $p_2$ are in-going, while momenta $p_3$ and $p_4$ are out-going. Here, $p$ denotes the centre-of-mass momentum, and $\chi$ is the scattering angle. The low-energy limit corresponds to the small transferred momentum limit $t\to 0$. This limit occurs when either $p\to 0$ or $\chi\to 0$. 

We begin with the case of quadratic coupling \eqref{the_quadratic_coupling_definition}. The matrix element describing the leading order corrections to the gravitational scattering is given by
\begin{align}
  \begin{split}
    \\
    &
    \begin{gathered}
      \begin{fmffile}{Light_Scattering_0}
        \begin{fmfgraph*}(30,30)
          \fmfleft{L1,L2}
          \fmfright{R1,R2}
          \fmf{dashes}{L1,V}
          \fmf{dashes}{L2,V}
          \fmf{dashes}{R1,V}
          \fmf{dashes}{R2,V}
          \fmfv{decoration.shape=circle,decoration.size=15pt,decoration.filled=shaded}{V}
          \fmflabel{$p_1$,$m\not=0$}{L1}
          \fmflabel{$p_3$,$m\not=0$}{L2}
          \fmflabel{$p_2$,$m=0$}{R1}
          \fmflabel{$p_4$,$m=0$}{R2}
        \end{fmfgraph*}
      \end{fmffile}
    \end{gathered}
    \hspace{50pt}= \mathcal{M}_\text{LO, quadratic}
    \\ \\
    &=
    \begin{gathered}
      \begin{fmffile}{Light_Scattering_1}
        \begin{fmfgraph}(40,40)
          \fmfbottom{B1,B2}
          \fmftop{T1,T2}
          \fmf{dashes}{B1,V1,T1}
          \fmf{dashes}{B2,V2,T2}
          \fmf{dbl_wiggly}{V1,V2}
          \fmfdot{V1,V2}
        \end{fmfgraph}
      \end{fmffile}
    \end{gathered}
    ~+~
    \begin{gathered}
      \begin{fmffile}{Light_Scattering_2}
        \begin{fmfgraph}(40,40)
          \fmfbottom{B2,B1}
          \fmftop{T2,T1}
          \fmf{dashes}{B1,V1,T1}
          \fmf{dashes}{B2,V2,T2}
          \fmf{dbl_wiggly}{V1,V}
          \fmf{phantom}{V,V2}
          \fmfdot{V1,V}
          \fmffreeze
          \fmf{dbl_wiggly,right=1}{V,V2,V}
          \fmfv{decoration.shape=circle,decoration.filled=empty,decoration.size=5}{V2}
        \end{fmfgraph}
      \end{fmffile}
    \end{gathered}
    ~+~
    \begin{gathered}
      \begin{fmffile}{Light_Scattering_3}
        \begin{fmfgraph}(40,40)
          \fmfbottom{B2,B1}
          \fmftop{T2,T1}
          \fmf{dashes}{B1,V1,T1}
          \fmf{dashes}{B2,V2,T2}
          \fmf{dbl_wiggly}{V1,V2}
          \fmfdot{V1}
          \fmfv{decoration.shape=circle,decoration.filled=empty,decoration.size=5}{V2}
          \fmffreeze
          \fmf{phantom,tension=2}{B2,V}
          \fmf{phantom,tension=0.1}{V,V2}
          \fmf{dbl_wiggly,left=1}{V2,V}
          \fmfdot{V}
        \end{fmfgraph}
      \end{fmffile}
    \end{gathered}
    ~+~
    \begin{gathered}
      \begin{fmffile}{Light_Scattering_4}
        \begin{fmfgraph}(40,40)
          \fmfbottom{B2,B1}
          \fmftop{T2,T1}
          \fmf{dashes}{B1,V1,T1}
          \fmf{dashes}{B2,V2,T2}
          \fmf{dbl_wiggly}{V1,V2}
          \fmfdot{V1}
          \fmfv{decoration.shape=circle,decoration.filled=empty,decoration.size=5}{V2}
          \fmffreeze
          \fmf{phantom,tension=2}{T2,V}
          \fmf{phantom,tension=0.1}{V,V2}
          \fmf{dbl_wiggly,right=1}{V2,V}
          \fmfdot{V}
        \end{fmfgraph}
      \end{fmffile}
    \end{gathered}
    ~+~
    \begin{gathered}
      \begin{fmffile}{Light_Scattering_5}
        \begin{fmfgraph}(40,40)
          \fmfbottom{B1,B2}
          \fmftop{T1,T2}
          \fmf{dashes,tension=2}{B1,V,T1}
          \fmfv{decoration.shape=circle,decoration.filled=empty,decoration.size=5}{V}
          \fmf{dbl_wiggly}{V,V1}
          \fmf{dbl_wiggly}{V,V2}
          \fmf{dashes,tension=2}{B2,V1}
          \fmf{dashes}{V1,V2}
          \fmf{dashes,tension=2}{V2,T2}
          \fmfdot{V1,V2}
        \end{fmfgraph}
      \end{fmffile}
    \end{gathered} .
  \end{split}
\end{align}
We have omitted the higher order terms in $\kappa$ and $\mathcal{E}$. The explicit matrix element expression is
\begin{align}
  \mathcal{M}_\text{LO,quadratic} = i\,\kappa^2\Bigg[ \cfrac{ ( M^2 - s ) ( M^2 - u ) }{4\, t} +\cfrac{7\, \pi^2 }{12\, \mathcal{E}^4 }\, M^2 \, t \,A_0 (M^2) - \cfrac{35\, \pi^2}{384\, \mathcal{E}^4}\,t^3\,B_0 (t,0,0) + \mathcal{O}\left( \frac{1}{\mathcal{E}^8} \right) \Bigg] .
\end{align}

The first term in this expression is relevant because it provides a dominant contribution in the $t\to 0$ limit. The second term is irrelevant since the $A_0$ integral does not depend on the transferred momentum, and the entire function vanishes in the $t \to 0$ limit. The last term has a singularity in the small momentum limit:
\begin{align}
  t^3 B_0 (t,0,0) = t^3 \left[ - \ln\left(-\frac{t}{\mu^2}\right) + \cfrac{1}{d-4} + 2 - \gamma - \ln\pi \right] \overset{t\to 0}{\sim} - t^3 \ln\left(-\frac{t}{\mu^2}\right) .
\end{align}
Here $1/(d-4)$ is the ultraviolet divergent factor, and $\gamma$ is the Euler constant. This function has a brunch cut at $t\geq 0$, so it is non-analytic at $t=0$ and contributes to the low energy limit. Therefore, the following part of the amplitude is leading in the low-energy limit:
\begin{align}
  \mathcal{M}_\text{LO,quadratic,IR} = i\,\pi\,G_\text{N}\,M^2\,\cfrac{8}{\sin^2\frac{\chi}{2}} \left[ 1 - \cfrac{70\,\pi^2}{3} \,\cfrac{ p^6 }{ \mathcal{E}^4\,M^2 }\,\sin^8\frac{\chi}{2}\,\ln\left( \cfrac{p^2}{\mu^2} \,\sin^2\frac{\chi}{2} \right) \right].
\end{align}
It provides the following contribution to the differential cross-section:
\begin{align}
  \left. \cfrac{d\sigma}{d\Omega} \right|_\text{LO,quadratic,IR} = \left(\cfrac{G_\text{N}\,M}{\sin^2\frac{\chi}{2}}\right)^2\left[1 - \cfrac{140\,\pi^2}{3}\,\cfrac{p^6}{\mathcal{E}^4\,M^2} ~ \sin^8\!\cfrac{\chi}{2} ~ \ln\left( \frac{p^2}{\mu^2}\,\sin^2\cfrac{\chi}{2} \right) +\, \mathcal{O}\left(\cfrac{1}{\mathcal{E}^6}\right)\right] + \mathcal{O}(G_\text{N}^3).
\end{align}
In turn, it admits the following small-angle expansion:
\begin{align}
  \begin{split}
    \left. \cfrac{d\sigma}{d\Omega} \right|_\text{LO,quadratic,IR} = (4\,G_\text{N}\,M)^2\Bigg[& \cfrac{1}{\chi^4} + \cfrac{1}{3!}\,\cfrac{1}{\chi^2} + \cfrac{11}{6!} + \cfrac{31}{6\times 7!}\, \chi^2 \\
      &+ \left\{ \cfrac{41}{2\times 9!} - \cfrac{35\,\pi^2}{8\times 4!} \, \cfrac{p^6}{\mathcal{E}^4\,M^2}\, \ln\left(\frac{p^2}{\mu^2} \sin^2\frac{\chi}{2}\right) \right\}\,\chi^4 + \mathcal{O}\left(\chi^5\right)\Bigg] .
  \end{split}
\end{align}

In complete agreement with the results of previous sections, the non-minimal coupling to the Gauss-Bonnet term does contribute to the scattering in the low energy limit. However, the contribution of this particular coupling is irrelevant for the small-angle scattering. Since the contribution appears at order $\mathcal{O}\left(\chi^4 \right)$, it vanishes in the small angle limit $\chi\to 0$ and cannot contribute to the small angle scattering. We return to this result in Section \ref{Section_conclusion}.

The situation is different for the case of the linear coupling \eqref{the_linear_coupling_definition}. The matrix element describing the scattering in this case reads:
\begin{align}
  \begin{split}
    \\
    \begin{gathered}
      \begin{fmffile}{Light_Scattering_0}
        \begin{fmfgraph*}(30,30)
          \fmfleft{L1,L2}
          \fmfright{R1,R2}
          \fmf{dashes}{L1,V}
          \fmf{dashes}{L2,V}
          \fmf{dashes}{R1,V}
          \fmf{dashes}{R2,V}
          \fmfv{decoration.shape=circle,decoration.size=15pt,decoration.filled=shaded}{V}
          \fmflabel{$p_1$,$m\not=0$}{L1}
          \fmflabel{$p_3$,$m\not=0$}{L2}
          \fmflabel{$p_2$,$m=0$}{R1}
          \fmflabel{$p_4$,$m=0$}{R2}
        \end{fmfgraph*}
      \end{fmffile}
    \end{gathered}
    \hspace{50pt} = \mathcal{M}_\text{LO,linear}=& ~
    \begin{gathered}
      \begin{fmffile}{Light_Scattering_1}
        \begin{fmfgraph}(40,40)
          \fmfbottom{B1,B2}
          \fmftop{T1,T2}
          \fmf{dashes}{B1,V1,T1}
          \fmf{dashes}{B2,V2,T2}
          \fmf{dbl_wiggly}{V1,V2}
          \fmfdot{V1,V2}
        \end{fmfgraph}
      \end{fmffile}
    \end{gathered}
    ~+~
    \begin{gathered}
      \begin{fmffile}{Light_Scattering_6}
        \begin{fmfgraph}(40,40)
          \fmfbottom{B2,B1}
          \fmftop{T2,T1}
          \fmf{dashes,tension=2}{B1,V1,T1}
          \fmf{dashes,tension=2}{B2,VD}
          \fmf{dbl_wiggly}{VD,VU}
          \fmf{dashes,tension=2}{VU,T2}
          \fmf{dbl_wiggly,tension=2}{V1,V}
          \fmf{dbl_wiggly}{V,VD}
          \fmf{dbl_wiggly}{V,VU}
          \fmfdot{V1,V}
          \fmfv{decoration.shape=circle,decoration.filled=empty,decoration.size=5}{VU}
          \fmfv{decoration.shape=circle,decoration.filled=empty,decoration.size=5}{VD}
        \end{fmfgraph}
      \end{fmffile}
    \end{gathered}
    \, . \\ \\ 
  \end{split} 
\end{align}
The explicit expression reads:
\begin{align}
  \begin{split}
    \mathcal{M}_\text{LO,linear} = i\,\kappa^2\Bigg[ & \cfrac{( M^2 - s) (M^2 - u)}{4\,t}   - \cfrac{\pi^2}{30720\,\mathcal{E}^6}\, \cfrac{1}{t\,\left( t-4\,M^2 \right)^2}\, B_0(t,0,0) \Bigg( 5760\, M^{14} + 960\, M^{12} (t-12\, u)\\
      & \hspace{10pt} - 32\, M^6\, t^2 \left(983\, t^2+720\, t \,u + 4\, u^2 \right) + 192\, M^{10} \left(t^2-20\, t\, u+30\, u^2\right) \\
      & \hspace{10pt} + 32\, M^8\, t \left(359\, t^2 + 158\, t \,u + 150\, u^2\right) + 16\, M^4\, t^3 \left(1299\, t^2 + 1504\, t\, u + 716\,  u^2\right) \\
      & \hspace{10pt} + t^5 \left(457\, t^2 + 832\, t\, u + 832\, u^2\right) - 2\, M^2\, t^4 \left( 2621\, t^2 + 3984\, t\, u + 3152\, u^2\right) \Bigg) \\
      & +\cfrac{\pi^2\,M^8}{32\,\mathcal{E}^6\,t\,(t-4\,M^2)^2}\,\Big( 6\, M^8 - 10\, M^4\, s\, u + s\, u\, (s+u)^2  \Big)\, C_0 (M^2 , M^2, t, 0,0,0) \\
      & - \cfrac{ \pi^2\, M^4}{63\,\mathcal{E}^6\,t\,(t - 4\,M^2)^2} \, B_0 (M^2,0 ,0) \Big( 52\, M^{10} - 2\, M^8 \, (33\, t+52\, u) \\
      & \hspace{10pt} + M^6\,  \left( 74\, t^2 + 168\, t\, u + 52\, u^2 \right) - M^4\, t \left( 43\, t^2 + 106\, t\, u + 58\, u^2 \right) \\
      & \hspace{10pt} + M^2\, t^2 \left(11\, t^2 + 30\, t\, u + 24\, u^2 \right) - t^3\, \left( t^2 + 3\, t\, u + 3\, u^2\right) \Bigg].
  \end{split}
\end{align}
In full analogy to the previous case, $B_0(M^2,0,0)$ is irrelevant in the low-energy limit since it is independent of the transferred momentum. The contribution of $B_0(t,0,0)$ is relevant because it is not analytic in $t\to 0$. The contribution of $C_0$ is much more sophisticated:
\begin{align}
  \begin{split}
    C_0 ( M^2\!, M^2\!, t, 0, 0, 0) &= -\cfrac{2}{ \sqrt{t (t \!-\! 4 M^2)} } \! \left[ \operatorname{Li}_2 \!\!\left(\!\! \cfrac{ \!-\! t \!+\! 2\,M^2 \!-\! \sqrt{t( t \!-\! 4 M^2)} }{2\, M^2} \right) \!-\! \operatorname{Li}_2 \!\! \left(\!\! \cfrac{ \!-\! t \!+\! 2\,M^2 \!+\! \sqrt{t ( t \!-\! 4 M^2)} }{2\, M^2} \right) \right]\\
    & \overset{t\to 0}{\to} -\cfrac{1}{M^2}\, \ln\left(-\cfrac{t}{M^2}\right).
  \end{split}
\end{align}
Here $\operatorname{Li}_2$ is the dilogarithm function.

In the low-energy limit, two different logarithms of $B_0$ and $C_0$ partially cancel each other out, giving the matrix element the following form:
\begin{align}
  \mathcal{M}_\text{LO,linear,IR} = i\,\pi^2\,G_\text{N}\,M^2 \,\cfrac{8}{\sin^2\frac{\chi}{2}}\left[ 1 -\cfrac{\pi^2}{128}\,\cfrac{M^6}{\mathcal{E}^6} ( 5 + \cos\chi) \ln\cfrac{M^2}{\mu^2} + \mathcal{O}\left(\cfrac{1}{\mathcal{E}^9 }\right)\right] + \mathcal{O}(G_\text{N}^3).
\end{align}
The following expression gives the corresponding expression for the scattering cross-section:
\begin{align}
  \left. \cfrac{d\sigma}{d\Omega} \right|_\text{LO,linear,IR} = \left(G_\text{N}\,M\,\cfrac{1}{\sin^2\frac{\chi}{2}}\right)^2\left[1 - \cfrac{\pi^2}{64}\,\cfrac{M^6}{\mathcal{E}^6} (5 + \cos\chi) \ln\cfrac{M^2}{\mu^2} +\, \mathcal{O}\left(\cfrac{1}{\mathcal{E}^9 }\right)\right] + \mathcal{O}(G_\text{N}^3).
\end{align}
It allows the following small-angle expansion:
\begin{align}\label{scattering_cross_section_linear_coupling}
  \begin{split}
    \left. \cfrac{d\sigma}{d\Omega} \right|_\text{LO,linear,IR} = (4\,G_\text{N}\,M)^2\Bigg[& \left\{ 1- \cfrac{3\,\pi^2}{32}\left(\frac{M}{\mathcal{E}}\right)^6 \ln\frac{M^2}{\mu^2} \right\} \cfrac{1}{\chi^4} + \left\{ 1- \cfrac{3\,\pi^2}{64}\left(\frac{M}{\mathcal{E}}\right)^6 \ln\frac{M^2}{\mu^2} \right\}\,\cfrac{1}{3!}\, \cfrac{1}{\chi^2}\\
      & +\left\{ 1- \cfrac{9\,\pi^2}{176}\left(\frac{M}{\mathcal{E}}\right)^6 \ln\frac{M^2}{\mu^2} \right\}\, \cfrac{11}{6!} +\mathcal{O}\left(\chi^1\right) \Bigg] .
  \end{split}
\end{align}

The non-minimal coupling does contribute to the small angle scattering for the linear coupling case. The contributions appear at order $\chi^{-4}$, $\chi^{-2}$, and $\chi^0$ and all contributions have different magnitude. This structure of corrections provides a way to test the theory prediction. Since the corrections are independent of the transferred momentum, the correction to $\chi^{-4}$ order can always be included in the observed mass $M_\text{obs}$:
\begin{align}
    \left( 4\, G_\text{N}\, M \right)^2 \left\{ 1- \cfrac{3\,\pi^2}{32}\left(\frac{M}{\mathcal{E}}\right)^6 \ln\frac{M^2}{\mu^2} \right\} = \left( 4\, G_\text{N}\, M_\text{obs} \right)^2 .
\end{align}
In other words, when studying the light scattering on a real object, one uses the $\chi^{-4}$ coefficient to define the mass of the central object. However, the definition of the observed mass does not eliminate the contributions to other orders in $\chi$:
\begin{align}
  \begin{split}
    \left. \cfrac{d\sigma}{d\Omega} \right|_\text{LO,linear,IR} = (4\,G_\text{N}\,M_\text{obs})^2\Bigg[& \cfrac{1}{\chi^4} + \left\{ 1 + \cfrac{3\,\pi^2}{64}\left(\frac{M}{\mathcal{E}}\right)^6 \ln\frac{M^2}{\mu^2} + \mathcal{O}\left(\mathcal{E}^{-7}\right) \right\}\,\cfrac{1}{3!}\, \cfrac{1}{\chi^2}\\
      & +\left\{ 1 + \cfrac{15\,\pi^2}{352}\left(\frac{M}{\mathcal{E}}\right)^6 \ln\frac{M^2}{\mu^2} + \mathcal{O}\left(\mathcal{E}^{-7}\right) \right\}\, \cfrac{11}{6!} +\mathcal{O}\left(\chi^1\right) \Bigg] .
  \end{split}
\end{align}
Consequently, in full accordance with the effective field theory, the linear coupling to the Gauss-Bonnet term contributes to the low energy limit. In contrast to the quadratic coupling, it contributes to the small-angle scattering at the $\chi^{-2}$ order. This allows one to search for opportunities to establish constraints on the new corrections with the empirical data on small-angle scattering. We discuss these results in the next section.

\section{Discussion and conclusion}\label{Section_conclusion}

This paper studied scalar-tensor gravity with non-minimal coupling to the Gauss-Bonnet term. These models are exciting since they evade the no-hair theorem and experience different dynamics.

Our study employs the perturbative quantum gravity, a practical approach that allows for the construction of an effective theory applicable below the Planck scale. While the theory is non-renormalisable, the need for complete renormalisation is negated by the effective nature of the theory.

The perturbative quantum gravity, in conjunction with the BPHZ theorem, provides a robust framework for studying gravity behaviour in the low energy limit. The BPHZ theorem, which establishes a general structure of a scattering amplitude within local quantum field theories, including perturbative quantum gravity, is a key theoretical tool in our research. In the low energy limit, the leading contribution to amplitude is given by non-analytic functions of external momenta, which are always free of UV divergent contributions. This means that in this limit, the contributions that depend on the renormalisation procedure are suppressed, while the renormalisation-independent contributions are dominant.

This approach was developed in a series of papers \cite{Donoghue:1994dn,Bjerrum-Bohr:2002gqz,Vanhove:2021zel,Travaglini:2022uwo,Bjerrum-Bohr:2022blt,Bjerrum-Bohr:2002fji}. It allows one to calculate the one-body non-relativistic gravitational potential and the leading order correction to an external gravitational field of a point-like quantum particle \cite{Donoghue:1994dn}. We apply this procedure to the discussed scalar-tensor models. In the case of linear coupling with the Gauss-Bonnet term, its contribution is suppressed by $r^{-3}$ in the non-relativistic potential \eqref{V_linear_coupling} and the metric \eqref{metric_linear_coupling}. In the case of quadratic coupling, its contribution is suppressed by $r^{-7}$ and $r^{-9}$ in the potential \eqref{V_quadratic_coupling} and the metric \eqref{metric_quadratic_coupling}. In both cases, we assume that the corresponding energy scales of the non-minimal coupling are below the Planck scale.

We studied the scattering of light-like particles described by perturbative quantum gravity and compared it to the light scattering of classical black holes within general relativity. Such scattering is crucial because it provides a way to analyse whether the new interaction can be found in experiments probing the motion of light around black holes. For the case of quadratic coupling, we have found that non-minimal Gauss-Bonnet coupling does not contribute to small-angle scattering, so it will only be relevant for large-angle scattering. On the contrary, the linear coupling to the Gauss-Bonnet term contributes to the small angle cross section \eqref{scattering_cross_section_linear_coupling}. The corresponding mass-scale hierarchy suppresses its contribution.

These results lead to the following general conclusions. A non-minimal coupling between a scalar field and the Gauss-Bonnet term affects gravitational physics in a highly non-trivial way. It introduces black holes with scalar hair into the theory and affects the structure of an external gravitational field of point-like particles. The corresponding changes in the external gravitational fields, probed by light scattering and the linear part of the non-minimal coupling, provide the leading order corrections. The forthcoming data on black hole shadows is expected to provide new empirical constraints on the theory.

\section*{Acknowledgement}

This work was supported by IBS, the Institute for Basic Science (Grant No. IBS-R018-Y1). We appreciate APCTP for its hospitality during the completion of this work.

\appendix

\section{Gauss-Bonnet term perturbative expansion}\label{Appendix_Gauss-Bonnet}

Below, we derive the leading order perturbative expansion of the Gauss-Bonnet term, which is given by the following formula:
\begin{align}
  \mathcal{G} = R^2 - 4 R_{\mu\nu}^2 + R_{\mu\nu\alpha\beta}^2.
\end{align}
Where $R_{\mu\nu\alpha\beta}$, $R_{\mu\nu}$, and $R$ represent the Riemann tensor, the Ricci tensor, and the scalar curvature, respectively. Within the perturbative approach, they admit the following perturbative expansions:
\begin{align}
  \begin{split}
    R_{\mu\nu\alpha\beta} &=-\cfrac{\kappa}{2}\left[ \pd_\mu\pd_\alpha h_{\nu\beta} - \pd_\mu\pd_\beta h_{\nu\alpha} + \pd_\nu\pd_\beta h_{\mu\alpha} - \pd_\nu \pd_\alpha h_{\mu\beta} \right] + \okappa{2},\\
    R_{\mu\nu} &=-\cfrac{\kappa}{2}\left[ \square h_{\mu\nu} + \pd_\mu\pd_\nu h - \pd_\mu\pd_\sigma h^\sigma{}_\nu - \pd_\nu \pd_\sigma h^\sigma{}_\mu \right] + \okappa{2} ,\\
    R &= - \kappa\left[ \square h - \pd_\mu\pd_\nu h^{\mu\nu} \right] + \okappa{2} .\\
  \end{split}
\end{align}
In the momentum representation, they correspond to the following expressions up to the leading order:
\begin{align}
  \begin{split}
    R_{\mu\nu\alpha\beta} \to & \cfrac{\kappa}{2} \left[ k_\mu k_\alpha I_{\nu\beta}{}^{\rho\sigma} - k_\mu k_\beta I_{\nu\alpha}{}^{\rho\sigma} + k_\nu k_\beta I_{\mu\alpha}{}^{\rho\sigma} -k_\nu k_\alpha I_{\mu\beta}{}^{\rho\sigma} \right] h_{\rho\sigma}(k) ,  \\
    R_{\mu\nu} \to& \cfrac{\kappa}{2} \left[ k^2 \,I_{\mu\nu}{}^{\rho\sigma} + k_\mu k_\nu \eta^{\rho\sigma} -\cfrac12\,\big( k_\mu k^\rho\delta_\nu{}^\sigma +k_\nu k^\rho \delta_\mu{}^\sigma+k_\mu k^\sigma\delta_\nu{}^\rho+k_\nu k^\sigma\delta_\mu{}^\rho \big) \right] h_{\rho\sigma}(k) , \\
    R \to&  \kappa \left[ k^2 \,\eta^{\rho\sigma} - k^\rho k^\sigma\right] h_{\rho\sigma}(k),
  \end{split}
\end{align}
where
\begin{align}
  I_{\mu\nu\alpha\beta} = \cfrac{1}{2} \left( \eta_{\mu\alpha}\eta_{\nu\beta} + \eta_{\mu\beta} \eta_{\nu\alpha} \right) .
\end{align}
Consequently, the leading contribution in the perturbative expansion of the Gauss-Bonnet term is of the order of $\okappa{2}$ and has the following structure:
\begin{align} 
  \mathcal{G} = \kappa^2\, \widehat{\mathcal{G}}^{\mu\nu\alpha\beta}(k_1,k_2) h_{\mu\nu}(k_1) h_{\alpha\beta}(k_2) + \okappa{2}. 
\end{align}
Here the tensor $\widehat{\mathcal{G}}_{\mu\nu\alpha\beta}$ has the following form
\begin{align}\label{Gauss-Bonnet_tensor_structure}
  \begin{split}
    \widehat{\mathcal{G}}_{\mu\nu\alpha\beta}(k_1,k_2) =& \cfrac14\, \eta_{\mu\nu}\eta_{\alpha\beta} \left[ k_1^2\,k_2^2 - 4\, (k_1\cdot k_2)^2 \right] + I_{\mu\nu\alpha\beta} \left[ (k_1\cdot k_2)^2 - k_1^2\,k_2^2 \right] \\
    &- \cfrac14\,\left[ \eta_{\mu\nu}\, k_1^2\, (k_2)_\alpha (k_2)_\beta + \eta_{\alpha\beta}\,k_2^2\,(k_1)_\mu (k_1)_\nu \right] - \left[ \eta_{\mu\nu}\,k_2^2\,(k_1)_\alpha(k_1)_\beta + \eta_{\alpha\beta}\,k_1^2\,(k_2)_\mu (k_2)_\nu \right]\\
    & + (k_1\cdot k_2) \left[ \eta_{\mu\nu} \left( (k_1)_\alpha (k_2)_\beta + (k_1)_\beta (k_2)_\alpha \right) + \eta_{\alpha\beta} \left( (k_1)_\mu (k_2)_\nu + (k_1)_\nu (k_2)_\mu \right) \right]\\
    & +\cfrac12\,\left\{ \eta_{\mu\alpha}\Big[k_1^2\,(k_2)_\nu (k_2)_\beta+k_2^2\,(k_1)_\nu (k_1)_\beta - (k_1\cdot k_2)\big((k_1)_\nu (k_2)_\beta + (k_1)_\beta (k_2)_\nu \big)\Big]+\cdots \right\}\\
    &+\cfrac14\,(k_1)_\mu (k_1)_\nu (k_2)_\alpha (k_2)_\beta + (k_1)_\alpha (k_1)_\beta (k_2)_\mu (k_2)_\nu - \cfrac12\left\{ (k_1)_\mu (k_1)_\alpha (k_2)_\nu (k_2)_\beta + \cdots \right\} .
  \end{split}
\end{align}
Dots $\cdots$ denote terms required by tensor symmetry:
\begin{align}
  \widehat{\mathcal{G}}_{\mu\nu\alpha\beta}(k_1,k_2) = \widehat{\mathcal{G}}_{\nu\mu\alpha\beta}(k_1,k_2) =\widehat{\mathcal{G}}_{\mu\nu\beta\alpha}(k_1,k_2) = \widehat{\mathcal{G}}_{\alpha\beta\mu\nu}(k_2,k_1) .
\end{align}  

\section{Fourier transformations}\label{Appendix_Fourier}

The present paper relies on Fourier transformations of non-analytic functions of a three-momentum. This appendix discusses a method to derive corresponding expressions and provide explicit formulas for the most critical cases.

The Fourier transform of a function that depends solely on the absolute value of a three-impulse can be simplified to a sin Fourier transform of a function of a single variable. This is achieved by successive integration over the angular degrees of freedom, which yields
\begin{align}
  \begin{split}
    &\int\cfrac{d^3\vec{p}}{(2\pi)^3} \,f \left(\abs{\vec{p}}\right) \, e^{-i\,\vec{p}\cdot\vec{r}} = \int\limits_0^\infty \, p^2\, \cfrac{d p}{2\pi} \int\limits_0^\pi \,\sin\theta \,\cfrac{d\theta}{2\pi}\int\limits_0^{2\pi} \cfrac{d\varphi}{2\pi} \, f(p) \, \exp\left[ - i\, p \, r \, \cos\theta\right] \\
    &= \cfrac{1}{2\pi^2}\,\cfrac{1}{r} \, \int\limits_0^\infty \, f(p)\,p\,\sin( p\, r) \, d p.
  \end{split}
\end{align}
We define the sin Fourier transform as follows:
\begin{align}
  \widehat{F}\left[ f(t) \right] (\omega) \overset{\text{def}}{=}  \sqrt{\cfrac{2}{\pi}}\,\int\limits_0^\infty f(t) \sin( \omega\, t) d t .
\end{align}
In these notations, the original three-dimensional integral can be expressed as a one-dimensional Fourier transform:
\begin{align}
  \int\cfrac{d^3\vec{p}}{(2\pi)^3} \,f \left(\abs{\vec{p}}\right) \, e^{-i\,\vec{p}\cdot\vec{r}} = \cfrac{1}{2\pi^2}\,\cfrac{1}{r}\,\sqrt{\cfrac{\pi}{2}}\,\widehat{F}\left[ f(p) \, p \right](r)
\end{align}

It is important to note that functions analytic in the $\vec{p}\to 0$ limit can correspond to powers of $1/r$. However, it seems unrealistic to expect such functions to appear in actual quantum field theory calculations:
\begin{align}
  \begin{split}
    \int\cfrac{d^3\vec{p}}{(2\pi)^3} \,p^{2n} e^{-i\vec{p}\cdot\vec{r}} =& \cfrac{(-1)^{n+1} \delta^{(2n+1)}(r)}{2\pi\,r}  \,, \quad n=0,1,\cdots ; \\
    \int\cfrac{d^3\vec{p}}{(2\pi)^3} \,p^{2n-1} e^{-i\vec{p}\cdot\vec{r}} =& \cfrac{(-1)^n\,(2n)!}{2\pi^2\, r^{2(n+1)}}\,, \quad n=1,2,\cdots.
  \end{split}
\end{align}

Non-analytic functions that grow as $\vec{p}\to 0$ can produce both positive and negative powers of $r$. Some of them also have logarithmic components:
\begin{align}
  \begin{split}
    \int\cfrac{d^3\vec{p}}{(2\pi)^3} \,\cfrac{1}{p^{2n}}\, e^{-i\vec{p}\cdot\vec{r}} =& \cfrac{(-1)^{n-1} \, r^{2(n-1)-1}}{4\pi\, (2\,n-2)!} \, , \quad n=1,2,\cdots ; \\
    \int\cfrac{d^3\vec{p}}{(2\pi)^3} \,\cfrac{1}{p^{2n+1}}\, e^{-i\vec{p}\cdot\vec{r}} =& \cfrac{(-1)^n\,r^{2(n-1)} \left(\ln(r) - \psi(2\,n)\right)}{2\pi^2\,(2n-1)!} \, , \quad n=1,2,\cdots . \\
  \end{split}
\end{align}
Here $\psi(z) \overset{\text{def}}{=} \Gamma'(z)/\Gamma(z)$ is the digamma function.

Finally, non-analytic functions containing $\ln p^2$ are of special interest. The logarithm itself admits a special point at $p=0$ where it diverges. The divergence is very weak since the function gets a finite limit start multiplied by any positive power of $p$:
\begin{align}
  \lim\limits_{p\to 0} \,p^n \, \ln p^2 =0 \,, n=1,2,\cdots.
\end{align}
At the same time, the function $p^n \ln p^2$ is not smooth enough in the limit $p\to 0$ and is not analytic at that point. Despite admitting finite limits, these functions produce negative powers of $r$:
\begin{align}
  \begin{split}
    \int\cfrac{d^3\vec{p}}{(2\pi)^3} ~ p^{2n}\,\ln p^2 \, e^{-i\,\vec{p}\cdot\vec{r}} &= \cfrac{(-1)^{n+1} \,(2n+1)! }{2\pi\,r^{2\,n+3} } \,, \quad n=0,1\cdots; \\
    \int\cfrac{d^3\vec{p}}{(2\pi)^3} ~ p^{2n-1}\,\ln p^2 \, e^{-i\,\vec{p}\cdot\vec{r}} &= \cfrac{(-1)^{n+1} \,(2n)! \Big( \ln r - \psi (2n+1) \Big) }{\pi^2\, r^{2(n+1)}} \,, \quad n=1,2\cdots . \\
  \end{split}
\end{align}

\bibliographystyle{unsrturl}
\bibliography{HairyS}

\end{document}